\documentclass[a4paper,11pt]{article}
\pdfoutput=1 % if your are submitting a pdflatex (i.e. if you have
\usepackage{jcappub} % for details on the use of the package, please
\usepackage[T1]{fontenc} % if needed

\usepackage{comment}
\usepackage{natbib}

\usepackage{hyperref} %[hidelinks,colorlinks=true,linkcolor=blue,citecolor=blue]

\def\reffig#1{Fig.~\ref{fig:#1}}

\title{\boldmath EFTofLSS meets simulation-based inference: $\sigma_8$ from biased tracers}

\author[a]{Beatriz Tucci,}
\author[a]{Fabian Schmidt}

\affiliation[a]{Max–Planck–Institut f\"ur Astrophysik, Karl–Schwarzschild–Stra\ss e 1, 85748 Garching, Germany}

\emailAdd{tucci@mpa-garching.mpg.de}
\emailAdd{fabians@mpa-garching.mpg.de}

\abstract{Cosmological inferences typically rely on explicit expressions for the likelihood and covariance of the data vector, which normally consists of a set of summary statistics. However, in the case of nonlinear large-scale structure, exact expressions for either likelihood or covariance are unknown, and even approximate expressions can become very cumbersome, depending on the scales and summary statistics considered. Simulation-based inference (SBI), in contrast, does not require an explicit form for the likelihood but only a prior and a simulator, thereby naturally circumventing these issues. In this paper, we explore how this technique can be used to infer $\sigma_8$ from a Lagrangian effective field theory (EFT) based forward model for biased tracers. The power spectrum and bispectrum are used as summary statistics to obtain the posterior of the cosmological, bias and noise parameters via neural density estimation. We compare full simulation-based inference with cases where the data vector is drawn from a Gaussian likelihood with sample and analytical covariances. We conclude that, for $k_{\text{max}}=0.1h\text{Mpc}^{-1}$ and $0.2h\text{Mpc}^{-1}$, the form of the covariance is more important than the non-Gaussianity of the likelihood, although this conclusion is expected to depend on the cosmological parameter inferred, the summary statistics considered and range of scales probed. }

\begin{document}
\maketitle
\flushbottom

\section{Introduction}
\label{sec:intro}

The current standard cosmological inference procedure consists in providing an analytical likelihood for the data vector considered, which together with parameter priors and sampling methods such as Monte-Carlo Markov Chain (MCMC) yields the posteriors of the parameters of interest given the observed data \cite{Trotta_2008, Verde_2010}. Focusing on the case of a galaxy catalog from a certain cosmological survey, it is standard to \textit{compress} the information contained in this catalog and infer the cosmological parameters using summary statistics as the data vector, such as the galaxy power spectrum \cite{Nishimichi_2020, dAmico_2020}.

The first natural question then is how to perform the galaxy field compression, in order to maximize the amount of information extracted. Many alternatives have been proposed in the literature to go beyond two-point statistics, such as higher $n$-point functions (bispectrum, trispectrum, etc), Minkowski functionals, wavelets, k-nearest neighbor cumulative distribution functions, peak counts, void statistics, among many others (see, e.g., \cite{Planck2015iso, Euclid_2023, Park_2023}). Of course, a given cosmological parameter, such as the amplitude of fluctuations $\mathcal{A}_s$ or dark energy equation of state $w$, affects different observables differently, so that the optimal compression choice is highly dependent on the parameter to be inferred. Recently, exciting developments are being made towards \textit{field-level} inference, where the information from the entire galaxy density field is used to sample both the cosmological parameters and the initial phases with Hamiltonian Monte Carlo (HMC) due to the high-dimensionality of the parameter space \cite{jasche2013bayesian, wang2014ELUCID, lavaux2019BOSS, kitaura2020cosmic, Schmidt_2020, Nguyen_2021, Babic_2022, Kostic_2022, Porqueres_2023}.

In this paper, we defer the difficult question of optimal compression, and instead focus on the lowest order $n$-point functions, namely the galaxy power spectrum and bispectrum. These summaries are well motivated on quasilinear scales by perturbative approaches to large-scale structure (LSS). Having fixed the definition of the data vector, the next problem consists in predicting the distribution of each of the summary statistics. For essentially all relevant LSS observables, only approximate expressions for these distributions are known. Even if an exact likelihood expression could be written down and used for inference, there would remain additional issues with this standard explicit-likelihood approach which we discuss in detail below; in particular, covariance estimation and binning effects. 

Simulation-based inference (SBI) \cite{Cranmer_2020, Lueckmann_2021}, also known as likelihood-free inference (LFI) or implicit-likelihood inference, arises as an alternative to this approach. SBI makes use of a \textit{simulator model} that specifies a process to generate ``mock'' data; that is, for a given set of parameters $\boldsymbol{\theta}$, the model generates (or forward-models) independent samples of $p(\textbf{x}|\boldsymbol{\theta})$, where $\textbf{x}$ is the data vector (of summary statistics, in our case). The inference makes use of the simulated dataset to obtain the posterior and, since it does not require a direct evaluation of the conditional density, is therefore in contrast to \textit{density models}, or explicit-likelihood models, where $p(\textbf{x}|\boldsymbol{\theta})$ has to be specified to obtain the posterior via Bayes' rule. This latter class of models is currently the standard in cosmology inference. 

Simulator models, by virtue of their construction, operate as dynamic representations of the actual mechanisms at work based on our well-established scientific knowledge, providing a direct and intuitive link between theoretical models and observable phenomena. They are therefore more natural and interpretable than density models, where the density function specification depends on a more abstract understanding of real-world phenomena and often requires analytical approximations \cite{Papamakarios_Thesis}.  

Although we work with $n$-point functions in this paper, it is important to stress that the SBI approach generalizes straightforwardly to any general data compression scheme, as long as the simulator still provides accurate predictions and the data vector is not extremely large. One can therefore study different summary statistics and their combinations with no concerns on finding an analytical expression for their distribution. This facilitates the use of more informative summary statistics, in particular those learned by neural networks \cite{Fluri_2021, Kacprzak_2022}, which can be flexibly optimized according to each particular inference problem. In the context of galaxy clustering, recent progress has been made by using graphs networks to capture the map information from the galaxy distribution \cite{Villanueva-Domingo_2022, Makinen_2022}.

SBI has been widely used in the cosmological inference context, including weak-lensing \cite{Alsing_2019a, Alsing_2019b, Jeffrey_2021, Lin_2022}, type IA supernovae \cite{Leclercq_2018, Alsing_2018, Alsing_2019a, Alsing_2019b, Karchev_2023, Bernardo_2023, Chen_2023}, standard sirens \cite{Gerardi_2021, Gagnon-Hartman_2023}, CMB \cite{Cole_2022, Lemos_2023b}, galaxy cluster abundance \cite{Reza_2022}, Gaussian and lognormal fields \cite{Leclercq_2021, Makinen_2021, Dai_2022, Akhmetzhanova_2023}, dark-matter overdensity fields \cite{Dai_2022, Modi_2023a}, voids \cite{Thiele_2023}, dark-matter halos \cite{Makinen_2022, Shao_2023} and galaxies \cite{Hahn_2022, Hahn_2023, deSanti_2023, Modi_2023b}. 

Nonetheless, no work has been done in the specific context of the Effective Field Theory of the Large Scale Structure (EFTofLSS) and the bias expansion \cite{Desjacques_2018}, which together provide a rigorous framework for modelling galaxy clustering as described in Section \ref{sec:LPT}. Recently, a significant effort has been made towards constraining cosmological parameters from galaxy clustering data using the usual likelihood-based analysis \cite{Nishimichi_2020, dAmico_2020, Ivanov_2020, Philcox_2020, DAmico_2022, Zhang_2022, Philcox_2022a, Philcox_2022b, Ivanov_2023}. Although this has been a successful program for cosmological inference, it is faced with significant practical and conceptual challenges, and we discuss below some of the main advantages of using SBI for galaxy clustering analysis instead.\\

\paragraph{Why simulation-based inference for galaxy clustering?}

\paragraph{Likelihood approximation.}

Focusing on the cosmological inference procedure for galaxy clustering, the use of an analytical approximation for the likelihood becomes already inadequate for the galaxy power spectrum, the ``simplest'' summary statistic. It is well known that the Gaussian approximation for cosmological two-point functions fails at low wavenumbers $k$, since they are estimated as a sum of field amplitudes squared, and cosmic variance breaks the central limit theorem due to the small number of modes \cite{2018MNRAS.473.2355S,2018MNRAS.477.4879S}.
This issue is investigated in \cite{2018MNRAS.477.4879S} for weak lensing power spectra, in \cite{2019MNRAS.485.2956H} for the 3D galaxy power spectrum, and in \cite{,2021MNRAS.506L..85L} for a log-normal density field. The Gaussian likelihood approximation particularly affects the posterior of parameters that are sensitive to low-$k$ modes, such as in the case of $f_{\text{NL}}$ \cite{Komatsu_2001}, which expresses deviations from Gaussianity in the perturbations generated during the inflation epoch. In addition, by coupling different Fourier modes, nonlinear evolution also breaks the Gaussian assumption for the likelihood on small scales, not only for the power spectrum but also for many other summary statistics.

\paragraph{Covariance estimation.} Even if the Gaussian likelihood assumption is accurate, we need to determine the covariance of the data vector.
The usual procedure is to either assume an analytical approximation, or to estimate the covariance from simulated mock catalogs with fixed cosmology. Analytical approximations are not always sufficient, and estimation from mocks might be cumbersome and unpractical depending on the summary statistics chosen. It is worth noting in this context that existing EFT-based analyses of galaxy clustering make use of covariances estimated from mocks, while the mean of the data vector is based on the EFT, which is strictly speaking inconsistent. In addition, there is the issue of at which (fixed or varying) point in parameter space the covariance of the data vector should be computed. SBI circumvents these issues and provides a means to use a consistent prediction of the entire distribution of the data vector and its dependence on the parameters $\boldsymbol{\theta}$, including correlations between different elements of the data vector and higher-order moments. All of this is of great importance for scientific reasoning and the assurance that the errors are not underestimated via a poor covariance estimation; however, specific testing methods have to be applied to the SBI posterior for an evaluation of whether it itself is underestimating the errors \cite{Hermans_2021}.

\paragraph{Binning effects.}
In standard explicit-likelihood approaches, it is necessary to bin-average the theory prediction for the data vector in accordance with the binning used for the data. This is know to be important in the bispectrum case, see for example \cite{Bernardeau_2012}. For the bispectrum, there is also the distinction between ``open'' and ``closed'' triangles, which arises from the fact that ``open'' triangle bins (i.e., those which do not satisfy the closed triangle condition) contain individual ``closed'' triangle modes \cite{Ivanov_2022}. These concerns are not present in the SBI approach, as the data vector can be freely chosen, as long as the same procedure is applied to observed and simulated data.

\paragraph{Forward modeling.} 

As already mentioned, SBI makes use of simulator models to obtain the simulated dataset and therefore is an application of forward modeling, where a prescription of the generative process for the data vector as a function of the model parameters is given. Although not particular to SBI, forward modeling presents several advantages over other methods, such as the straightforward inclusion of observational effects, including window functions, redshift-space distortions and systematic effects \cite{2018MNRAS.473.2355S,2019A&A...624A.115P,2023MNRAS.520.6283Y,Hahn_2023,2023arXiv230103581T,Stadler_2023}. In the context of the EFT forward model employed here, forward modeling at the field-level allows us to reach essentially arbitrary perturbative (loop) order \cite{Schmidt_2021}.

\paragraph{}

To sum up, galaxy clustering for cosmological precision analysis is inevitably moving towards the need of higher $n$-point functions, combining different summary statistics and expanding the range of scales probed. It is evident how the covariance computation complexity scales with higher $n$-point functions, and how the need of an analytical approximation for the likelihood limits the summary statistics which can be used for analysis. Given a trustable simulator model, SBI therefore provides a flexible and rigorous framework for statistical inference in the context of galaxy clustering,  provided that the following points are addressed: \textit{model misspecification}, or how accurate is the simulator, \textit{convergence}, or how many simulations are needed for a reliable posterior estimation, and \textit{diagnostics}, or how reliable is the estimated posterior. In particular, we are interested in determining the \textit{calibration} of the posterior \cite{Talts_2018, Hermans_2021} to check whether the uncertainties of the posterior are at least not underestimated.

In this paper, we take advantage of our understanding of galaxy clustering on large scales to test the SBI performance based on \texttt{LEFTfield}, a rigorous simulator based on the EFTofLSS. Using this tool as a first step is of crucial importance, since we tackle model misspecification by keeping all modes under control, cutting small scales where we should not trust our simulator and using the bias expansion on the scales where it is known to be robust. The simulator uses Lagrangian Perturbation Theory (LPT), which is fast to evaluate and therefore allows a detailed study of convergence with respect to the number of simulations and also careful posterior diagnostics. 

Using the bias expansion constructed in \texttt{LEFTfield}, we employ the combination of the galaxy power spectrum and bispectrum to break the degeneracy between the bias parameters and the amplitude of fluctuations $\sigma_8$. Our main goal is to assess the potential impact on cosmological inference when we consider the entire data vector distribution through SBI rather than relying on the assumption of a Gaussian likelihood. We also provide a comprehensive description of how simulation-based inference works in practice, as this is one of the first applications of this technique to galaxy clustering. In the context of SBI, this is the first time that the complete second-order bias expansion is used. Our work is thus complementary to recent work \cite{Hahn_2022, Hahn_2023, Modi_2023b}, which uses halo occupation distribution (HOD) as a bias model, and includes the power spectrum on smaller scales. Note that our forward model is numerically much less costly than the one used in \cite{Hahn_2022, Hahn_2023, Modi_2023b}, allowing us to generate many more independent simulated data realizations.

The paper is structured as follows. In Section \ref{sec:LPT}, we review perturbation theory for galaxy clustering, while the details on the forward model can be found in Section \ref{sec:LEFTfield}. The power-spectrum and bispectrum estimators used are described in Section \ref{sec:summaries}, and we explain in detail how simulation-based inference works in Sections \ref{sec:ABC} and \ref{sec:NDE}. The results are shown in Section \ref{sec:linear_bias} (linear forward model), and in Section \ref{sec:2LPT2d} (LPT-based forward model). We conclude in Section \ref{sec:conclusion}. 

\section{Galaxy clustering}

\subsection{Perturbation Theory}
\label{sec:LPT}
In order to study the dynamics of matter, we can write the geodesic equation for the Eulerian comoving position $\boldsymbol{x}(\tau)$ of a Newtonian pressureless fluid element and relate it to the gravitational potential $\Phi(\boldsymbol{x},\tau)$. If we denote the Lagrangian position of this fluid at $\tau=0$ as $\boldsymbol{q}$, we obtain
\begin{equation}
    \boldsymbol{x}(\boldsymbol{q}, \tau) = \boldsymbol{q} + \boldsymbol{s}(\boldsymbol{q}, \tau),
    \label{eq:displacement}
\end{equation}
where $\boldsymbol{s}(\boldsymbol{q}, \tau)$ is the Lagrangian displacement along the fluid trajectory, which is our dynamic variable in the Lagrangian picture. We can then write the geodesic equation for the displacement and relate it to the gravitational potential $\Phi(\boldsymbol{x},\tau)$, which in turn is connected to the matter density contrast $\delta(\boldsymbol{x},\tau)$ via the Poisson equation. The continuity equation for matter combined with Eq. \eqref{eq:displacement} further implies that
\begin{equation}
    1+\delta(\boldsymbol{x}, \tau) = |\boldsymbol{1}+\boldsymbol{M}(\boldsymbol{q},\tau)|^{-1},
    \label{eq:delta_M}
\end{equation}
where we defined the Lagrangian deformation tensor $M_{ij} \equiv \partial s_j/\partial{q_i}$, allowing us to write the equation of motion of the displacement in terms of $\boldsymbol{q}$ only \cite{Matsubara_2015}. 

Lagrangian Perturbation Theory (LPT) then proceeds by expanding the displacement as \cite{Buchert_1992}
\begin{equation}
    \boldsymbol{s}(\boldsymbol{q},\tau)=\sum_{n=1}^\infty \boldsymbol{s}^{(n)}(\boldsymbol{q},\tau)
    \label{eq:displacement_expansion}
\end{equation}
and equivalently for the deformation tensor to obtain the LPT recursive relations \cite{Rampf_2012b, Zheligovsky_2014, Matsubara_2015}. It is also convenient to separate the displacement into a longitudinal $\sigma\equiv\boldsymbol{\nabla}\cdot\boldsymbol{s}$ and a transverse component $\boldsymbol{t}\equiv\boldsymbol{\nabla}\times\boldsymbol{s}$, 
\begin{equation}
    \boldsymbol{s} = \frac{\boldsymbol{\nabla}}{\nabla^2}\sigma - \frac{1}{\nabla^2}\boldsymbol{\nabla}\times\boldsymbol{t},
    \label{eq:displacement_decomposition}
\end{equation}
where the resulting evolution equations show that the transverse contribution $\boldsymbol{t}$ only becomes relevant at third order \cite{Rampf_2012a, Schmidt_2021}. The time component can be factorized out, which allows for writing evolution equations for $\sigma$ and $\boldsymbol{t}$ order by order as a function of their initial conditions, and also for the construction of a closed bias expansion (see below). Although this factorization is mostly used in the context of an Einstein deSitter (EdS) Universe, it is also possible for a general expansion history \cite{Ehlers_1997}.

The equivalence principle guarantees that the leading observable for non-relativistic observers is the tidal field $\partial_i\partial_j\Phi$ and its time derivatives along the fluid trajectory. In the EFTofLSS context \cite{Baumann_2012, Carroll_2014, Senatore_2015, Abolhasani_2016, Porto_2016}, we aim to describe the number density of galaxies by including all possible dependencies on this quantity at a given fixed order in perturbation theory. By doing so, we refrain ourselves from modeling all unknown small-scale physics, as they can be absorbed into EFT parameters to be determined from data. The ultraviolet (or small-scale) cutoff $\Lambda$ should be seen as a computational tool that guarantees loop integrals remain finite, as any observable has to be independent of its choice. In the following, we will not explicitly indicate the cutoff $\Lambda$ for clarity, but re-instate it in the next section.

Since at leading order in spatial derivatives all the allowed terms by EFTofLSS are contained in $\boldsymbol{M}(\boldsymbol{q},\tau)$ and its time derivatives \cite{Mirbabayi_2015}, the deterministic part of the galaxy overdensity field\footnote{We will be using the term ``galaxies'' throughout the paper, since the bias expansion is valid for any biased tracer.} in the Lagrangian picture can be written as
\begin{equation}
    \delta^L_{g, \text{det}}(\boldsymbol{q}, \tau) = \int_0^\tau d\tau' F_g\left[\boldsymbol{M}(\boldsymbol{q},\tau'),\tau',\tau\right],
\end{equation}
where $F_g$ is a nonlocal in time functional. By expanding $F_g$ in $\boldsymbol{M}$, we obtain all the rotational invariants of $\boldsymbol{M}$, and the time integral can be performed thanks to the aforementioned time factorization. We can construct the Lagrangian bias expansion as

\begin{equation}
    \delta^L_{g, \text{det}}(\boldsymbol{q}, \tau) = \sum_{\mathcal{O}^L} b_{\mathcal{O}^L}(\tau) \mathcal{O}^L(\boldsymbol{q}, \tau),
    \label{eq:Lagrangian_bias_expansion}
\end{equation}
where $b_{\mathcal{O}^L}$ denote the Lagrangian bias parameters that encode all complex and non-linear physics of galaxy formation from $F_g$. The Lagrangian operators $\mathcal{O}^L$ are determined by taking all scalar invariants of the symmetric part of $\boldsymbol{M}^{(n)}$ and their independent products. It is not necessary to include $\text{tr}[\boldsymbol{M}^{(n)}]$ for $n>1$ nor the antisymmetric contributions, since they can be expressed in terms of lower-order symmetric ones via the equations of motion \cite{Zheligovsky_2014, Matsubara_2015}. The local basis of the deterministic Lagrangian operators up to second order is then given by \cite{Mirbabayi_2015}
\begin{equation}
    \begin{split}
        1^{\text{st}}\quad \quad &\text{tr}[\boldsymbol{M}^{(1)}]\\
        2^{\text{nd}}\quad \quad &\text{tr}[\boldsymbol{M}^{(1)}\boldsymbol{M}^{(1)}], \,(\text{tr}[\boldsymbol{M}^{(1)}])^{2}.
    \end{split}
    \label{eq:Lagrangian_basis}
\end{equation}

So far, we have assumed that the galaxy formation process is perfectly local in space. Going beyond this approximation requires the introduction of higher-(spatial-)derivative contributions, the leading of which is $\nabla^2\delta$. By dimensional analysis, these contributions are associated with a spatial length scale $R_*$. Supposing that $R_*$ is of the same order as the nonlinearity scale, then within the perturbative regime it is sufficient to keep the leading order high-derivative term $b_{\nabla^2\delta}\nabla^2\delta(\boldsymbol{x}, \tau)$, as the higher-order ones are suppressed by more powers of $(R_*k)^2$.  The higher-derivative bias are naturally introduced when imposing that the observables should not depend on the EFT cutoff \cite{Desjacques_2018}, where it is important to stress that the nonlocality scale $R_*$ differs from the cutoff scale $\Lambda^{-1}$. In particular, all the counterterms needed to renormalize the operators of Eq. \eqref{eq:Lagrangian_basis} are the ones already listed there, together with their associated higher-order derivatives terms \cite{Mirbabayi_2015}. 

The observed galaxy density field is in Eulerian space, while our expansion so far has been treated in Lagrangian space. We use Eq. \eqref{eq:displacement} to \textit{advect} the Lagrangian operators to Eulerian space. The continuity equation for galaxies gives the relation between the galaxy overdensity in Eulerian and Lagrangian coordinates, i.e.,
\begin{equation}
    \delta_{g,\text{det}}(\boldsymbol{k},\tau) = \int \frac{d\boldsymbol{q}}{(2\pi)^3}  \,
    e^{i\boldsymbol{k} \cdot \left[\boldsymbol{q} + \boldsymbol{s}(\boldsymbol{q},\tau)\right]}
    \left[1+\delta_g^L(\boldsymbol{q},\tau)\right].
    \label{eq:eulerian_lagrangian_relation}
\end{equation}
By expanding the displacement as in Eq. \eqref{eq:displacement_expansion} and the exponential in Eq. \eqref{eq:eulerian_lagrangian_relation}, it is possible to obtain the relation between the Lagrangian and Eulerian basis order by order \cite{Abidi_2018}. It is worth emphasizing that Lagrangian and Eulerian descriptions correspond merely to different coordinate choices, that is, one does not assume a conserved galaxy number. In the Eulerian case, we perform a perturbative expansion from Eq. \eqref{eq:delta_M} in terms of the matter overdensity instead of the Lagrangian displacement. 

So far we have discussed the deterministic operators, as in the form of Eq. \eqref{eq:Lagrangian_bias_expansion}, but the complete basis in the bias expansion should also account for the stochastic effect of small scales on galaxy formation. Since the stochastic fields are uncorrelated with large-scale perturbations, they are completely described by their $n$-point functions. For the two- and three point functions, we have that, at leading order and at a fixed time slice,
\begin{equation}
    \begin{split}
        \big\langle\delta_g(\boldsymbol{k}_1)\delta_g(\boldsymbol{k}_2)\big\rangle^{\prime} \Big|_{\text{stoch.}}^{\text{LO}} &= P_\varepsilon,\\
        \big\langle\delta_g(\boldsymbol{k}_1)\delta_g(\boldsymbol{k}_2)\delta_g(\boldsymbol{k}_3)\big\rangle^{\prime}\Big|_{\text{stoch.}}^{\text{LO}} &= B_\varepsilon + 2\,b_1P_{\varepsilon\varepsilon_\delta}\,\big(P_{\text{m}}(\boldsymbol{k}_1)+2\text{ }\text{perm.}\big),
    \end{split}
    \label{eq:LO_stoch_spectra}
\end{equation}
where the prime denotes that the momentum-conserving Dirac delta function is dropped, $b_1$ is the Eulerian bias associated to $\delta(\boldsymbol{x},\tau)$ and $P_m$ is the non-linear matter power spectrum. The noise parameters $P_\varepsilon$, $B_\varepsilon$ and $P_{\varepsilon\varepsilon_\delta}$ are to be determined together with the bias parameters, where the posterior of the cosmological parameters can be obtained by marginalizing over them. 

Just like for large-scale perturbations, galaxy formation also depends nonlocally on small-scale perturbations inside the region characterized by $R_*$, such that we also have to take into account higher-derivative contributions for the stochastic fields. In Fourier space, these contributions are suppressed by $k^2 R_*^2$ on large scales \cite{Desjacques_2018}, and we neglect them here. 

In our EFT forward model we construct all operators starting from the linear, sharp-$k$-filtered density field contrast $\delta^{(1)}_\Lambda$. This is related to the expansion of $\boldsymbol{M}$ by expanding Eq. \eqref{eq:delta_M} to first order, yielding
\begin{equation}
    \sigma^{(1)}_\Lambda \equiv \text{tr}[\boldsymbol{M}_\Lambda^{(1)}] =
    - \delta^{(1)}_\Lambda.
    \label{eq:sigma_linear_density}
\end{equation}
The higher-order LPT basis terms are then constructed out of $\sigma^{(1)}_\Lambda$ via the LPT recursion relations. We stress the difference between our finite, explicit cutoff $\Lambda$ and the formal cutoff used in semi-analytical loop calculations, which is usually sent to $\infty$ \cite{Carroll_2014, Assassi_2014, Rubira:2023vzw}. 

\subsection{LEFTfield}
\label{sec:LEFTfield}

The forward model we use to get from the initial conditions, cosmological, noise and bias parameters to the final data is \texttt{LEFTfield}, which evolves a given initial (linear) density field up to \textit{n}-th order in LPT \cite{Schmidt_2021}. This represents a powerful tool for cosmological inference using SBI. As already discussed, statistical inference in SBI is dependent on the quality of the simulator and the simulation budget. Using the EFT-based approach, we keep all the modes under control up to a certain scale $\Lambda$, such that SBI only learns the features we trust from our simulator. We therefore combine two big advantages of EFTofLSS for SBI: (i) theoretically robust treatment of bias; and (ii) fast evaluation thanks to the fact that we only need to follow the evolution of modes up to $\Lambda$, making it a perfectly suitable tool for analysing convergence and coverage of SBI algorithms. For example, using a Subhalo Abundance Matching (SHAM) \cite{Vale_2004} approach would be less theoretically robust, while being much more costly in comparison. The price to pay of course is that our forward model is limited to scales where perturbation theory is valid.

Our model differs from those commonly used in the EFTofLSS literature where the basis for the $n$-point functions are generated by methods such as CLASS-PT \cite{Chudaykin_2020} and PyBird \cite{DAmico_2020b}, since we directly measure the $n$-point functions on the simulated galaxy density field.  By doing so, we are also simulating the full (non-Gaussian) distribution of our data vector, as generated by the LPT-based simulator, and SBI can provide us with the correct distribution for the summary statistics considered. Each simulation realization is run with a different seed for the density and stochastic fields, so that our model correctly accounts for cosmic variance. Note that SBI models the \emph{probability density} of the data vector, and is therefore also in contrast to emulators \cite{Angulo_2021}, where the \emph{expectation value} of the summary statistics is learned. Also, our simulator naturally incorporates IR resummation by way of the Lagrangian forward model.

\texttt{LEFTfield}, which is based on \cite{Schmidt_2021}, works as follows. We start from an initial power spectrum $P_L$ which is scaled by $\alpha^2$, where we define $\alpha\equiv\sigma_8/\sigma_8^{\text{fid}}$ and set $\sigma_8^{\text{fid}}=0.85$, while all other cosmological parameters are fixed to $\Omega_m=0.3$, $\Omega_\Lambda=0.7$, $h=0.7$ and $n_s=0.967$. A Gaussian random field $\delta^{(1)}_\Lambda$ of size $N_g^3$  is then generated from this scaled power spectrum as
\begin{equation}
    \delta^{(1)}_\Lambda(\boldsymbol{k},z) = W_\Lambda(k) \sqrt{\alpha^2 P_L(k,z)}\,\hat{s}(\boldsymbol{k}),
    \label{eq:delta_ini}
\end{equation}
where $\hat{s}(\boldsymbol{k})$ is a Gaussian random field of zero mean and unit variance, $z$ is the redshift and the grid is smoothed at scale $\Lambda$ with a sharp-$k$ filter $W_\Lambda(k)$, guaranteeing that we are only treating perturbative modes in the forward model \cite{Schmidt_2020}. We emphasize that the inferred bias and stochastic parameters will depend on (``run with'') $\Lambda$ \cite{Rubira:2023vzw}, while inferred cosmological parameters such as $\alpha$ should be independent of $\Lambda$. We then use the relation from Eq. \eqref{eq:sigma_linear_density} to set our initial conditions for $\sigma^{(1)}_\Lambda$, while $\boldsymbol{M}$ and $\boldsymbol{t}$ are constructed in the same grid using the evolution equations. The displacement $\boldsymbol{s}(\boldsymbol{k})$ is then evaluated from Eq. \eqref{eq:displacement_decomposition} up to the desired order in LPT. 

We then use the displacement to advect the Lagrangian operators to Eulerian space. As previously discussed, in perturbation theory this is done by both expanding the exponential and the displacement in Eq. \eqref{eq:eulerian_lagrangian_relation}. With the aim of preventing noise generation on large-scales, instead of expanding the exponential we use a cloud-in-cell (CIC) scheme, where the ``mass of the particles'' corresponds to the value of the Lagrangian operators at each grid cell, i.e., $M_{\text{CIC}}^{\mathcal{O}}(\boldsymbol{q})=\mathcal{O}^L(\boldsymbol{q})$, while the extra $1+\delta$ term from the Jacobian is absorbed into the bias parameters. This approach is similar to what has been previously done in the literature; however, while \cite{Schmittfull_2019} uses the Zel'dovich approximation (1LPT), we expand the displacement up to $n$-th order and \cite{Kokron_2021, Zennaro_2021} estimate the full displacement from N-body simulations. In this paper, we use second-order LPT (2LPT) throughout.

After the displacement $\boldsymbol{s}(\boldsymbol{k})$ is inversely Fourier transformed to configuration space and copied to a larger grid of dimension $N_{\text{CIC}}^3$, we use Eq. \eqref{eq:displacement} to advect the deterministic Lagrangian operators $O^L$ listed in Eq. \eqref{eq:Lagrangian_basis}, with the exception of $\text{tr}[\boldsymbol{M}^{(1)}]$. Since $\delta$ and $\boldsymbol{M}$ are nonlinearly related through Eq. \eqref{eq:delta_M}, the usual bias term $b_1\delta(\boldsymbol{q})$ contributes to several terms of Eq. \eqref{eq:Lagrangian_basis}, so we instead displace a field with mass $M_{\text{CIC}}^{\delta}=1$ to obtain the Eulerian density directly. Although displacing $\text{tr}[\boldsymbol{M}^{(1)}]$ would equivalently lead to a complete basis of operators, this approach allows us to work with the linear bias $b_1$, facilitating the interpretation of our results. The deterministic bias expansion in Eulerian space then reads
\begin{equation}
    \begin{split}
        \delta_{g,\text{det}}(\boldsymbol{x}, \tau) &= b_1(\tau)\delta(\boldsymbol{x}, \tau) 
        + b_{\nabla^2\delta}(\tau)\nabla^2\delta(\boldsymbol{x}, \tau)\\
        &+ b_{\text{tr}[M^{(1)} M^{(1)}]}(\tau)\,\text{tr}[\boldsymbol{M}_\Lambda^{(1)}\boldsymbol{M}_\Lambda^{(1)}](\boldsymbol{x}, \tau) 
        + b_{\sigma\sigma}(\tau)\big[\sigma^{(1)}_\Lambda\big]^2(\boldsymbol{x}, \tau),
    \end{split}
    \label{eq:LEFTfield_deltag}
\end{equation}
where $\big[\sigma^{(1)}_\Lambda\big]^2 = (\text{tr}[\boldsymbol{M}_\Lambda^{(1)}])^2$, and we have omitted the subtraction of the mean of the second-order operators for clarity. The stochastic contributions are added directly in Eulerian space, such that the final galaxy overdensity field reads
\begin{equation}
    \delta_{g}(\boldsymbol{x}, \tau) = \delta_{g, \text{det}}(\boldsymbol{x}, \tau) + \varepsilon(\boldsymbol{x}, \tau)+ c_{\varepsilon\delta}(\tau)\varepsilon(\boldsymbol{x}, \tau)\delta(\boldsymbol{x}, \tau) + c_{\varepsilon^2}(\tau) \varepsilon^2(\boldsymbol{x}, \tau),
    \label{eq:deltag}
\end{equation}
where we also omitted the mean subtraction for $\varepsilon^2$. This expression exhibits some differences from the usual stochastic contributions in the bias expansion \cite{Desjacques_2018}. Since the simulated stochastic field $\varepsilon$ is sampled from a Gaussian distribution with zero mean and variance $P_\varepsilon$, non-Gaussianity of the noise, specifically the noise bispectrum $B_\varepsilon$ is generated via the $\varepsilon^2$ term in the model. Precisely, we obtain
\begin{equation}
    \big\langle\delta_g(\boldsymbol{k}_1)\delta_g(\boldsymbol{k}_2)\delta_g(\boldsymbol{k}_3)\big\rangle^{\prime} \Big|_{\text{stoch.}}^{\text{LO}} = 6\,c_{\varepsilon^2}P_\varepsilon^2 + 2\,b_1c_{\varepsilon\delta}P_{\varepsilon}\,\big(P_{m}(\boldsymbol{k}_1)+2\text{ }\text{perm.}\big),
\end{equation}
which thus captures both the $B_\varepsilon$ and $P_{\varepsilon\varepsilon_\delta}$ terms in Eq. \eqref{eq:LO_stoch_spectra} via $c_\varepsilon^2$ and $c_{\varepsilon\delta}$, as derived in Appendix \ref{sec:LO_noise_spectra}. In the Poisson limit, i.e., when the stochasticity of galaxies perfectly follows Poisson statistics, determined by their mean comoving density $\bar{n}_g$, we expect that $c_{\varepsilon^2}=1/6$ and $c_{\varepsilon\delta}=b_1/2$ (Appendix \ref{sec:LO_noise_spectra}). For a flowchart representation of how our forward model is constructed, we refer the reader to \cite{Schmidt_2021, Kostic_2022}.

\subsection{Summary statistics}
\label{sec:summaries}

The usual procedure to extract information from the galaxy density field is through its $n$-point functions. Since its one-point function (the mean density) is trivial, most of the information can be captured by its two- and three-point functions or, equivalently, their Fourier transforms, namely the power spectrum and bispectrum. Under the homogeneity and isotropy conditions, the bispectrum is described by three parameters which characterize the shape and the scale of the triangles, thus encoding important cosmological information beyond the two-point function which only depends on scale \cite{Sefusatti_2006}. Especially in the context of primordial non-Gaussianities, it is convenient to separate the triangle shapes into e.g. ``squeezed'' and ``equilateral'' configurations, but we do not make such distinctions as we work with all triangle configurations.

The tree-level galaxy power spectrum displays a degeneracy between the amplitude of density fluctuations, parametrized by $\alpha$ in our forward model, and the linear bias $b_1$, while the bispectrum has the power of breaking this degeneracy \cite{Fry_1994,Frieman_1994,Matarrese_1997,Sefusatti_2006}. The intuition is that, at leading order in perturbations, the galaxy bispectrum will depend on a sum of different powers of the bias parameters and the linear power spectrum, allowing for a determination of the linear and second-order bias parameters which is independent of the power spectrum normalization. We refer the reader to \cite{Jeong_2009} and \cite{Desjacques_2018} (Sec. 4.1.1) for an illustration of the different shape contributions of the matter and galaxy bispectrum, respectively.

To measure such $n$-point functions on the grids generated by \texttt{LEFTfield}, we use the bispectrum estimator introduced by \cite{Scoccimarro_1998}. We start by defining the quantities \cite{Gualdi_2021}
\begin{equation}
    I_k(\boldsymbol{x}) = \int_{||\boldsymbol{p}|-k|<\frac{\Delta k}{2}} \frac{d\boldsymbol{p}}{(2\pi)^3}\delta_g(\boldsymbol{p})e^{i\boldsymbol{x}\cdot\boldsymbol{p}}, \quad 
    J_k(\boldsymbol{x}) = \int_{||\boldsymbol{p}|-k|<\frac{\Delta k}{2}} \frac{d\boldsymbol{p}}{(2\pi)^3}e^{i\boldsymbol{x}\cdot\boldsymbol{p}},
\end{equation}
where all $\boldsymbol{p}$ modes within a $k$-shell of width $\Delta k$ centered on $k$ are integrated. The expressions for $I_k(\boldsymbol{x})$ and $J_k(\boldsymbol{x})$ therefore represent the inverse Fourier transform over a $k$-shell of the galaxy overdensity field, which is calculated following the procedure of Sec. \ref{sec:LEFTfield}, and a unit field, respectively. The power spectrum estimator is defined as 
\begin{equation}
    \hat{P}(k_1) = \left(\frac{L^3}{N^6}\right) \frac{\sum_{i=1}^{N^3} I^D_{k_1}(x_i) I^D_{k_1}(x_i)}{\sum_{j=1}^{N^3} J^D_{k_1}(x_j) J^D_{k_1}(x_j)},
\end{equation}
where the superscript $D$ refers to the discretized version of the fields, $k_1$ is the Fourier bin, $N$ is the grid size and $L$ is the box size, while the bispectrum estimator in turn is
\begin{equation}
    \hat{B}(k_1,k_2,k_3) = \left(\frac{L^6}{N^9}\right) \frac{\sum_{i=1}^{N^3} I^D_{k_1}(x_i) I^D_{k_2}(x_i) I^D_{k_3}(x_i)}{\sum_{j=1}^{N^3} J^D_{k_1}(x_j) J^D_{k_2}(x_j) J^D_{k_3}(x_j)}.
\end{equation}
It is now evident that the power spectrum comes essentially for free when estimating the bispectrum. As already discussed in the introduction, we do not suffer from binning effects, since bin averaging is automatically taken into account in the SBI framework. This allows us to use both ``open'' and ``closed'' triangle configurations, as opposed to likelihood-based approaches where the binning effect is important for the open triangles \cite{Oddo_2020, Ivanov_2022}. ``Open'' triangles are those for which the $k$-bins do not respect the momentum conservation constraint $|k_3-k_2|<k_1<k_2+k_3$, although there are individual triples of modes associated to the bin which do satisfy the relation.

Considering the most costly model in this work, i.e., the Euclid configuration described in Section \ref{sec:alpha_free}, using \texttt{LEFTfield} to generate $N_{\text{sim}}=10^5$ density grids with 2LPT, construct a second-order bias expansion and compute their corresponding power spectra and bispectra with data vector size of $D=49$ (6 bins for the power spectrum and 43 triangle bins for the bispectrum) takes roughly 19500 core hours when using 2GHz Intel Xeon Gold 6138 CPUs.

\begin{figure}
    \centering
    \includegraphics[width=\textwidth]{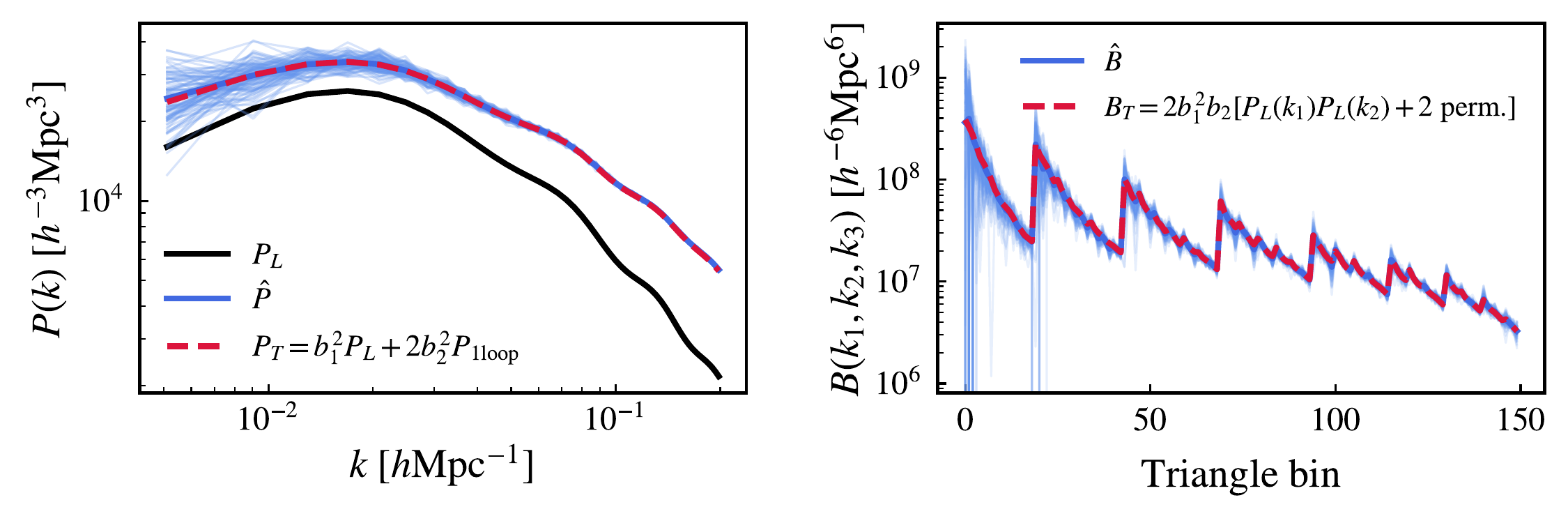}
    \caption{Comparison of power spectrum and bispectrum estimators to theory for the linear forward model in Eq. \eqref{eq:fwd_trivial_b2}, where we use $\Lambda=0.3h\text{Mpc}^{-1}$, $k_{\text{min}}=2\pi L^{-1} h\text{Mpc}^{-1}$ and $k_{\text{max}}=0.2h\text{Mpc}^{-1}$ for both cases. The blue lines show the $n$-point function estimator from \texttt{LEFTfield} as described in this section, which agree with the theory predictions indicated by the red curves. \textit{Left:} we set $b_1=b_2=1$ and use 50 linear bins for the power spectrum. The linear power spectrum $P_L$ is displayed in black for reference. \textit{Right:} we set $b_1=1$ and $b_2=0.1$, in order to suppress next-to-leading-order corrections, and use 150 triangle bins for a better visualization of the bispectrum, which are constructed from 10 linear bins in $k$. The lowest triangle bin indexes contain the smaller $k$-modes, hence their large variance, where $k_1 \le k_2 \le k_3$, where $k_1$ corresponds to the outermost loop index, while $k_3$ is the innermost index.}
    \label{fig:spectra_estimators}
\end{figure}

We show in Figure \ref{fig:spectra_estimators} the comparison of our estimators to their respective theoretical predictions. We use here a linear forward model instead of 2LPT, which is given directly from Eq. \eqref{eq:delta_ini} for $\alpha=1$, where we construct the final density field in a box with size $L=2000h\text{Mpc}^{-1}$ as 
\begin{equation}
    \delta_g(\boldsymbol{k},z) = b_1 \, \delta^{(1)}(\boldsymbol{k},z) + b_2 \, \left( [\delta^{(1)}]^2(\boldsymbol{k},z)-\langle[\delta^{(1)}]^2\rangle(z)\right),
    \label{eq:fwd_trivial_b2}
\end{equation}
choosing the redshift $z=0$. The second-order bias term generates a nonzero bispectrum.
The prediction for the power spectrum estimator is then $P_T(k)=b_1^2P_L(k)+2b_2^2P_{\text{1loop}}(k)$, where
\begin{equation}
    P_{\text{1loop}}(k) = \int_{||\boldsymbol{p}|-k|<\frac{\Delta k}{2}} \frac{d\textbf{p}}{(2\pi)^3} P_L\left(\left|\textbf{p}\right|\right) P_L\left(\left|\textbf{k}-\textbf{p}\right|\right) e^{i\textbf{k}\cdot\textbf{p}}
\end{equation}
and $P_L$ has support up to $\Lambda$. For the bispectrum, its leading-order theory prediction in this case reads $B_T(k_1, k_2, k_3) = 2b_1^2b_2 [P_L(k_1)P_L(k_2) + 2\,\, \mathrm{perm.}]$. We have also tested the LPT predictions for the power spectrum and bispectrum estimator (not shown), where in the latter case we compare the closed triangle bins only. 

\section{Simulation-based inference}
\label{sec:SBI}

We describe in detail how simulation-based algorithms work in this section. Its fundamental requirement is to have a \textit{simulator model}, such as \texttt{LEFTfield}. Let $\boldsymbol{\theta}$ be the vector of parameters of interest of dimension $N_\theta$ (e.g., cosmological, bias and noise parameters) and $\textbf{x}$ be the data vector of dimension $D$ (e.g., power-spectrum bins). Assuming that we have a proposal distribution over the parameters $\tilde{p}(\boldsymbol{\theta})$ (not necessarily the prior) and a simulator that generates $\textbf{x}$ given $\boldsymbol{\theta}$, it is possible to generate a simulated dataset of $N_{\text{sim}}$ samples $\{(\boldsymbol{\theta}_n, \textbf{x}_n)\}_{n=1}^{N_{\text{sim}}}$, where $(\boldsymbol{\theta}_n, \textbf{x}_n)$ is a joint sample from $p(\boldsymbol{\theta}, \textbf{x})=p(\textbf{x}|\boldsymbol{\theta})\tilde{p}(\boldsymbol{\theta})$. We describe below how the methods which will be used throughout this paper estimate the parameters posterior from this dataset given the observed data $\textbf{x}_o$.

\subsection{Approximate Bayesian Computation}
\label{sec:ABC}

The first approach to perform SBI goes back to the idea of Approximate Bayesian Computation (ABC). In its simplest form, rejection ABC \cite{Pritchard_1999}, one defines a distance metric $\rho(\textbf{x}, \textbf{x}_o)$ between the simulated samples $\textbf{x}$ and the observed data $\textbf{x}_o$, and it is possible to obtain samples from an approximate posterior $p(\boldsymbol{\theta}|\rho(\textbf{x}, \textbf{x}_o)<\epsilon)$ by accepting the parameters $\boldsymbol{\theta}_n$ for which the corresponding metric $\rho(\textbf{x}_n, \textbf{x}_o)$ lies below a given threshold $\epsilon$. In other words, the posterior is obtained by the histogram of parameters for which the simulated samples most closely resembles the observed data. In the limit $\epsilon\rightarrow0$ and under suitable regularity conditions, the approximated posterior recovers the true posterior \cite{Prangle_2015}. 

In practice, however, the value of $\epsilon$ controls the sample acceptance rate and the required simulation budget $N_\text{sim}$, so due to computational efficiency reasons it can't be chosen to be arbitrarily small. This is in tension with the aim for the smallest possible threshold to obtain a better approximation of the posterior. Extensions of this method were proposed to improve its efficiency such as Population Monte Carlo (PMC) \cite{Akeret_2015}, where a sequential approach is used and the approximated posterior is used as proposal distribution $\tilde{p}(\boldsymbol{\theta})$ for the next round until it has converged. These methods however still discard all the simulations which were rejected. Below, we discuss more efficient algorithms in which the entire dataset is used for the posterior estimation.

\subsection{Neural density estimators}
\label{sec:NDE}

One of the biggest problems related to rejection-based algortihms such as ABC and PMC concerns the fact that these inherently have to throw away all the simulations which lie below a certain threshold. Neural Density Estimators (NDEs) come as a rescue, since they use all available simulations to estimate the target distribution. The dataset of parameters and simulated summaries $\{(\boldsymbol{\theta}_n, \textbf{x}_n)\}_{n=1}^{N_{\text{sim}}}$ is used to find a certain probability distribution. Neural Posterior Estimation (NPE) \cite{Greenberg_2019} and Neural Likelihood Estimation (NLE) \cite{Papamakarios_2018} can use \textit{normalizing flows} \cite{Papamakarios_2019} to fit the posterior or the likelihood, respectively. There are also methods to learn the likelihood-to-evidence ratio which use classifiers instead, such as Neural Ratio Estimation (NRE) \cite{Hermans_2019, Durkan_2020, Miller_2021, Miller_2022, Delaunoy_2022}, but we do not discuss them here as they are not used for our analysis. We give a detailed explanation of normalizing flows in Appendix \ref{sec:normalizing_flows}.

\paragraph{Neural Likelihood Estimator.}

Suppose that we aim to approximate the unknown target distribution $p(\textbf{x}|\boldsymbol{\theta})$ as the conditional of the data \textbf{x} on the parameters $\boldsymbol{\theta}$. The idea is to fit a flow-based model $q_\phi(\textbf{x}|\boldsymbol{\theta})$ parametrized by $\phi$ to the dataset $\{(\boldsymbol{\theta}_n, \textbf{x}_n)\}_{n=1}^{N_{\text{sim}}}$ by imposing that the model approximates the target distribution, where $\tilde{p}(\boldsymbol{\theta},\textbf{x})=p(\textbf{x}|\boldsymbol{\theta})\tilde{p}(\boldsymbol{\theta})$. Evidently we do not know the target, but we can proceed by using the maximum likelihood estimation method which minimizes the forward KL divergence in the support of the proposals,
\begin{equation}
\begin{split}
    \mathbb{E}_{\tilde{p}(\boldsymbol{\theta})} \big[D_{\text{KL}}\,[\,p(\textbf{x}|\boldsymbol{\theta})\,||\,q_\phi(\textbf{x}|\boldsymbol{\theta})\,] \big] 
    & = \int d\boldsymbol{\theta} \, \tilde{p}(\boldsymbol{\theta}) \int d\textbf{x} \,
    p(\textbf{x}|\boldsymbol{\theta}) \, \log \left(\frac{p(\textbf{x}|\boldsymbol{\theta})}{q_\phi(\textbf{x}|\boldsymbol{\theta})}\right) \\
    & = \int d\boldsymbol{\theta} \, d\textbf{x} \,
    \tilde{p}(\boldsymbol{\theta},\textbf{x}) \, \log \left(\frac{p(\textbf{x}|\boldsymbol{\theta})}{q_\phi(\textbf{x}|\boldsymbol{\theta})}\right) \\
    & = -\mathbb{E}_{\tilde{p}(\boldsymbol{\theta},\textbf{x})}[\,\log q_\phi(\textbf{x}|\boldsymbol{\theta})\,] \, + \, \text{const.} \\
    & \approx -\frac{1}{N_{\text{sim}}} \sum_{n=1}^{N_{\text{sim}}} \log q_\phi(\textbf{x}_n|\boldsymbol{\theta}_n) \, + \, \text{const.}\,,
\end{split}
\label{eq:KL_divergence}
\end{equation}
where on the third line we identified the part which is independent of $\phi$ as a constant and on the last line we approximated the expectation over $\tilde{p}(\boldsymbol{\theta},\textbf{x})$ with Monte Carlo. The resulting Monte Carlo estimate of the KL divergence we wish to minimize is therefore independent of the explicit form of the target distribution and equivalent to the sum over the negative log-likelihood of the simulated datasets batches under the flow-based model. 

The estimation of conditionals is a natural extension of Masked Autorregressive Flows (MAF) \cite{Papamakarios_2017}, where only the conditionals corresponding to $\textbf{x}$ are modelled by augmenting the set of input variables with $\boldsymbol{\theta}$, i.e., $\textbf{z}_{t}=f_t(\textbf{z}_{t-1},\boldsymbol{\theta})$ (see Appendix \ref{sec:normalizing_flows}). The only requirement is that $\textbf{x}$ appears before $\boldsymbol{\theta}$ for any order used, and no connections have to be masked out from $\boldsymbol{\theta}$ to the rest of the network. Therefore, the Neural Likelihood Estimator (NLE) builds the flow-based model with MAF as \cite{Papamakarios_2018}
\begin{equation}
    q_\phi(\textbf{x}|\boldsymbol{\theta}) = \mathcal{N}(\textbf{z}_0|\mathbf{0}, \mathbf{I})\prod_{t=1}^T \left| \text{det} \left( \frac{\partial f_t}{\partial \textbf{z}_{t-1}} \right) \right|^{-1},
\end{equation}
where the network parameters are trained with the loss given by Eq. \eqref{eq:KL_divergence}. 

Note that here the model is trained for \textit{any given data} $\textbf{x}$, where for inference we are interested in a specific observed data $\textbf{x}_o$. The procedure is then to use the likelihood as the estimated conditional evaluated at the observed data, and then use MCMC (which is very cheap since only the learned model has to be evaluated) together with a prior $p(\boldsymbol{\theta})$ to obtain the estimated posterior $\hat{p}(\boldsymbol{\theta}|\textbf{x}_o)\propto q_\phi(\textbf{x}_o|\boldsymbol{\theta})p(\boldsymbol{\theta})$.

When simulating the data is expensive, it is often desirable to use \textit{sequential} approaches such as in Sequential Neural Likelihood Estimator (SNLE) \cite{Papamakarios_2018}, where the density estimation is done over multiple \textit{rounds} focusing on a given observation $\bar{\textbf{x}}$. The first proposal $\tilde{p}(\boldsymbol{\theta})$ is often chosen to be the prior, while the subsequent proposals are set as the estimated posterior $\hat{p}(\boldsymbol{\theta}|\bar{\textbf{x}})$ from the previous round. This of course limits the density estimation to a single observation and is therefore \textit{non-amortized}, but has the advantage of requiring less simulations to obtain an accurate posterior estimation focused on $\textbf{x}_o$, as the regions where the proposal density is high tend to be better approximated by the model. However, the cost of calibrating non-amortized methods is huge, as discussed below.

\paragraph{Neural Posterior Estimator.}

An advantage of the SNLE method is that the prior can be changed and tested, but it has the MCMC as an extra computational step. This can be avoided by estimating the posterior directly, as in the case of Sequential Neural Posterior Estimator (SNPE) \cite{Papamakarios_2016, Lueckmann_2017, Greenberg_2019}. The estimated posterior has the form
\begin{equation}
    \hat{p}(\boldsymbol{\theta}|\textbf{x})=p(\boldsymbol{\theta}|\textbf{x})\frac{\tilde{p}(\boldsymbol{\theta}) p(\textbf{x})}{p(\boldsymbol{\theta})\tilde{p}(\textbf{x})},
\end{equation}
where $\tilde{p}(\textbf{x})=\int_{\boldsymbol{\theta}}\tilde{p}(\boldsymbol{\theta})\tilde{p}(\textbf{x}|
\boldsymbol{\theta})$. Note that the \textit{proposal posterior} $\hat{p}(\boldsymbol{\theta}|\textbf{x})$ is equal to the target posterior if the proposal $\tilde{p}(\boldsymbol{\theta})$ equals the prior $p(\boldsymbol{\theta})$. 

In its first implementation \cite{Papamakarios_2016}, the undesired dependence on the proposal $\tilde{p}(\boldsymbol{\theta})$ was adjusted by multiplying the estimated posterior $\hat{p}(\boldsymbol{\theta}|\textbf{x})$ by $p(\boldsymbol{\theta})/\tilde{p}(\boldsymbol{\theta})$. This restricts $q_\phi(\boldsymbol{\theta}|\textbf{x})$ to be a Mixture Gaussian Density \cite{Bishop_1994} and $\tilde{p}(\boldsymbol{\theta})$ to be a Gaussian, so that the division can be done analytically; additionally, $p(\boldsymbol{\theta})$ can only be Gaussian or uniform. Besides these strong requirements, a drawback of this model is the fact that the division can lead to non-positive covariance matrices if $\tilde{p}(\boldsymbol{\theta})$ has smaller variance than any of the components of $q_\phi(\boldsymbol{\theta}|\textbf{x})$.

A further extension to this method was presented in \cite{Lueckmann_2017}, where the aforementioned requirements are no longer needed and the method no longer yields negative covariances. This is achieved by weighting the samples with $w_n=p(\boldsymbol{\theta}_n)/\tilde{p}(\boldsymbol{\theta}_n)$, such that the loss is modified to $-\sum_n w_n \log  q_\phi(\textbf{x}_n|\boldsymbol{\theta}_n)$. However, the weights can have high variance, leading to instability during the training process and inaccurate predictions in some cases \cite{Greenberg_2019}. The algorithm developed in \cite{Greenberg_2019} circumvents all these issues with the idea of ``atomic'' proposals, that essentially replaces integrals by sums and allows for the usage of arbitrary density estimators, especially those based on normalizing flows such as MAFs.

\paragraph{}

We have tested some of the NDE algorithms implemented in the publicly available \texttt{SBI} package \cite{Tejero-Cantero_2020}, and we chose to work with NPE from \texttt{SBI} as our baseline. There are several reasons for our choice; first, this method has been empirically observed to perform better for the inference problems considered in this work, especially for a limited simulation budget and large dimensions of the data parameter vectors. Second, it does not require an extra MCMC step as in NLE and NRE, or a retraining of the model for each observation such as for non-amortized algorithms, allowing for faster posterior diagnostics. Lastly, amortized posteriors such as the ones obtained by NPE tend to be more conservative \cite{Hermans_2021}. 

We use the SNPE method of \cite{Greenberg_2019} with 10 atoms for atomic proposals and MAF with 5 autoregressive layers (i.e., stacked MADEs), each constructed using two fully-connected tanh layers with 50 hidden units. We train the models by stochastically minimizing the loss using the Adam optimizer \cite{Kingma_2014} with learning rate of $5\times10^{-4}$ and batch size of 50, where 10\% of the samples are used for validation and the training is stopped if the validation set loss did not improve after 20 consecutive epochs. 

Training the most complex case considered here (Section \ref{sec:alpha_free}, with data vector size of $D=49$, $N_\theta=8$ parameters and simulation budget of $N_{\text{sim}}=10^5$) takes roughly one hour on a single 2GHz Intel Xeon Gold 6138 CPU, and no significant improvement was observed when using a GPU. Generating $10^5$ samples from the estimated posterior takes less than one second, allowing for efficient calibration analysis since the NPE posterior is amortized.

\begin{figure}
    \centering
    \includegraphics[width=\textwidth]{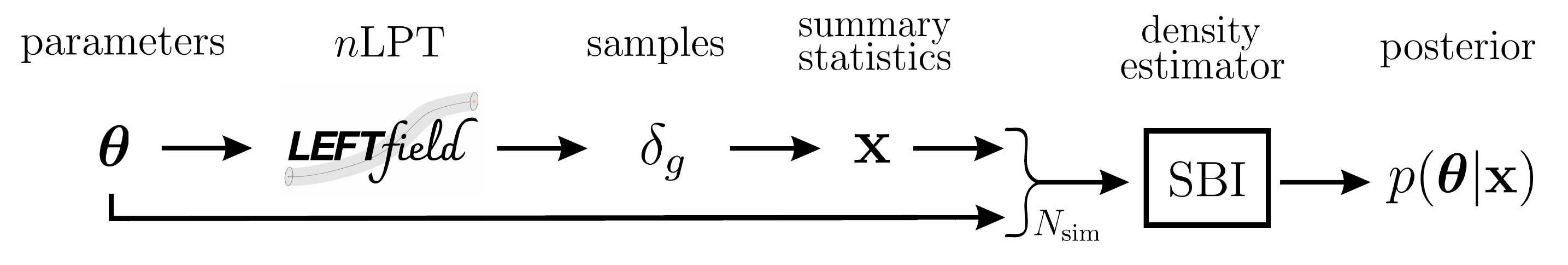}
    \caption{Flowchart of our pipeline for SBI, assuming an $n$LPT forward model. The cosmological, bias and noise parameters $\boldsymbol{\theta}$ are sampled from the prior and used as input for \texttt{LEFTfield} (Sec. \ref{sec:LEFTfield}), which constructs a model for the galaxy overdensity field $\delta_g$ through $n$LPT (2LPT in this work). We then measure the summary statistics $\textbf{x}$, which in our case is the galaxy power-spectrum concatenated with the bispectrum (Sec. \ref{sec:summaries}). We repeat this process $N_{\text{sim}}$ times, and use the resulting pairs $\{(\boldsymbol{\theta}_n, \textbf{x}_n)\}_{n=1}^{N_{\text{sim}}}$ as input to the SBI neural density estimation (Sec. \ref{sec:NDE}), which then gives us the posterior $p(\boldsymbol{\theta}|\textbf{x})$ that can in turn be evaluated at a given observed data $\textbf{x}_o$.}
    \label{fig:sbi_diagram}
\end{figure}

\section{Results}
\label{sec:results}

In this section, we present our results regarding two cases: first, a simple, linear forward model, described in Sec. \ref{sec:linear_bias}, where we compare ABC and NPE. In Sec. \ref{sec:2LPT2d}, we then use a 2LPT forward model and estimate the posterior with NPE, as summarized in Figure \ref{fig:sbi_diagram}.

\subsection{Linear forward model}
\label{sec:linear_bias}

We start reporting our results with a very simple model, where we aim to infer the linear bias $b_1$ and the noise amplitude $P_\varepsilon$ from the galaxy power spectrum. For that,  we consider a \textit{linear forward model}, where instead of using LPT we use Eq. \eqref{eq:delta_ini} with $\alpha=1$ to construct the galaxy overdensity field as
\begin{equation}
    \delta_g(\boldsymbol{k},z) = b_1 \, \delta^{(1)}(\boldsymbol{k},z) + \varepsilon(\boldsymbol{k}),
    \label{eq:fwd_linearbias}
\end{equation}
where $\varepsilon(\boldsymbol{k})$ is a Gaussian random field of zero mean and variance $P_{\varepsilon}$. We use a box of size $L=2000\,h^{-1}$Mpc, and choose redshift $z=0$, without any loss of generality in this case.

We define the observed data $\textbf{x}_o$ as a fiducial power spectrum $\bar{P}(k)$ measured from a particular realization of the model evaluated at the ground-truth parameters $\bar{b}_1=1.5$ and $\bar{P}_{\varepsilon}=10^3 \, h^{-3}\text{Mpc}^3$. We then generate $N_{\text{sim}}$ simulated datasets, where $\boldsymbol{\theta}=\{b_1^2, \, P_\varepsilon\}$ are sampled from a Gaussian prior as $\theta\sim\mathcal{N}(\bar{\theta}, \, 10^{-1}\times\bar{\theta})$ and $\textbf{x}=\{P(k)\}=\{P(k_1), P(k_2),\ldots,P(k_{D})\}$, where $D$ is the number of power-spectrum bins $N_{\text{bin}}$, are the power spectra of the simulated datasets. It is important to stress that, for each simulation, the realization of both $\delta^{(1)}(\boldsymbol{k})$ and $\varepsilon(\boldsymbol{k})$ are also varied. We infer $b_1^2$ instead of $b_1$ since its posterior can be evaluated analytically assuming a Gaussian likelihood and prior. Note that the SBI posterior for $b_1^2$ does not necessarily converge to the analytical one, since we relax the Gaussian likelihood assumption in this case.

The ABC metric we consider is
\begin{equation}
    \rho(\textbf{x}, \textbf{x}_o) = \frac{1}{(D-N_{\theta}-1)}\sum_{i=1}^{D} \frac{(\textbf{x}^i-\textbf{x}_o^i)^2}{\text{Cov}[\bar{\textbf{x}}^i]},
\end{equation}
where the sum runs over the summary statistics bins ($k$-bins for the power spectrum). This metric is inspired by the reduced $\chi^2$, and is expected to be close to optimal in the present case, in addition to being straightforward to interpret. Here, $D=N_{\text{bin}}=4$, $N_\theta=2$ and $\text{Cov}[\bar{P}(k)]=2\langle\bar{P}(k)\rangle^2/N_k$, where $\langle\bar{P}(k)\rangle$ is the mean of the fiducial power spectrum over initial conditions evaluated at fiducial parameters $\bar{\boldsymbol{\theta}}$ and $N_k$ is the total number of modes which lie in the respective $k$-bin. As explained in Sec. \ref{sec:ABC}, the proposed parameters $\boldsymbol{\theta}$ are then accepted or rejected according to this metric evaluated at the simulated data vector $\textbf{x}$, and the ABC posterior should converge to the true one in the $\epsilon\rightarrow0$ limit. However, due to computational efforts, one can never truly reach this limit.

\begin{figure}
    \centering
    \includegraphics[width=0.75\textwidth]{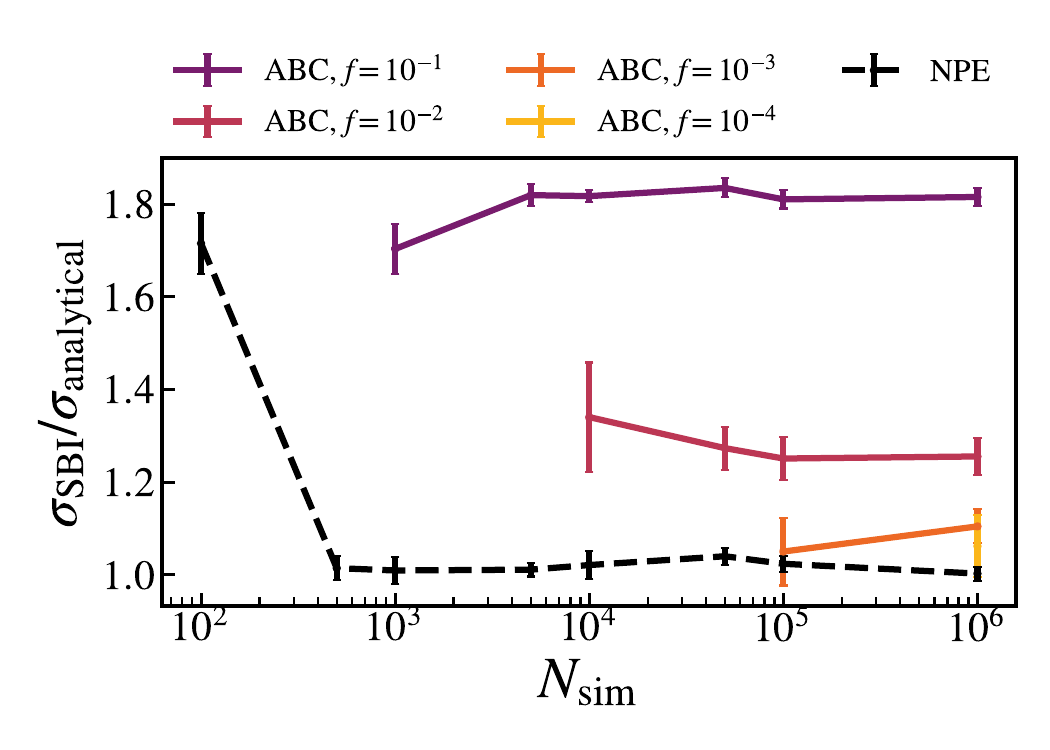}
    \caption{Posterior standard deviation of the linear bias parameter squared $b_1^2$ obtained by SBI divided by its analytical prediction as a function of the total simulation budget  $N_{\text{sim}}$. Continuous lines show the results obtained from ABC, while the dashed line corresponds to NPE. The value of $f$ denotes the fractional percentage of samples which were accepted in ABC, where we use $f \times N_{\text{sim}}$ for posterior estimation. The error bars correspond to 10 different realizations of the observational mock data $\textbf{x}_o$.}
    \label{fig:spectra_abc}
\end{figure}

We take the approach of \cite{Lueckmann_2021}, where instead of defining a specific threshold $\epsilon$ we accept the parameters corresponding to the lowest quantile $f$ of metric values. We show in Figure \ref{fig:spectra_abc} the SBI posterior standard deviation for $b_1^2$ normalized by the analytical prediction, where we compare ABC to NPE. We use a minimum of $10^2$ samples for estimating the posterior variance, leading to a different number of points for different choices of $f$.  
The corresponding threshold value $\epsilon$ for each percentage fraction $f=\{10^{-1},\,10^{-2},\,10^{-3},\,10^{-4}\}$ is $\epsilon=\{ 16.2, \,3.6, \,1.1, \,0.4\}$.

First, we can see that the NPE posterior converges much faster in terms of $N_{\rm sim}$ than the ABC one, as expected since the effective number of samples used in ABC is $f \times N_{\text{sim}}$, although the total simulation cost is $N_{\text{sim}}$, while NPE uses all simulations at hand for posterior estimation. Further, we find that one has to be careful when choosing the value for the treshold $\epsilon$; values that are too large will lead to significantly overestimated error bars, regardless of the number of simulations used. If the summary statistics used is approximately Gaussian-distributed with approximately known covariance, such as in our case, defining the ABC metric via the reduced $\chi^2$ makes the former interpretable, with convergence expected for values of $\epsilon$ of order unity, which is indeed confirmed by Figure~\ref{fig:spectra_abc} given the values of $\epsilon$ listed above. 
In the simple case we considered here, we can see that the posterior standard deviation converges to the Gaussian prediction, and we can conclude that for this particular case the non-Gaussianity of the power spectrum at low-$k$ does not significantly impact the constraints. 

Figure~\ref{fig:spectra_abc} serves as a useful cross-check of the different SBI approaches. In the following, due to the difficulty in achieving converged posteriors via ABC, we will use NPE throughout.

\subsection{2LPT forward model}
\label{sec:2LPT2d}

We now move to a forward model consisting of second-order LPT and second-order bias expansion as described in Section \ref{sec:LEFTfield}, where the galaxy overdensity field is given by Eq. \eqref{eq:deltag}. Here, $\textbf{x}$ is the concatenated vector of power spectrum and bispectrum bins of dimension $D=N_{\text{bin}}+N_{\text{tr}}$, where $N_{\text{tr}}$ denotes the number of triangle bins for the bispectrum. We use the configuration of the Euclid satellite \cite{Amendola_2013}, which will cover a volume of $63\,h^{-3}$Gpc$^3$ with mean number density of galaxies $\bar{n}_g=5.2\times10^{-4}\,h^{-3}$Mpc$^3$ at mean redshift $z=1.4$.

We consider two cutoffs for generating the datasets. For $\Lambda=0.1 \, h\text{Mpc}^{-1}$, the grid size is $N_c=128^3$ and the dimension of the data vector is $D=33$, while for $\Lambda=0.2 \, h\text{Mpc}^{-1}$, $N_c=256^3$ and $D=49$. The fiducial values of the bias and noise parameters are set to
\begin{equation}
\begin{split}
    \bar{\boldsymbol{\theta}} &=  \left\{\bar{\alpha}, \, \bar{b}_1, \, \bar{b}_{\nabla^2\delta}, \, \bar{b}_{\text{tr}[M^{(1)}M^{(1)}]}, \, \bar{b}_{\sigma \sigma}, \, 10^{-3}\bar{P}_\varepsilon, \, \bar{c}_{\varepsilon\delta}, \, \bar{c}_{\varepsilon^2}  \right\} \\
    &= \left\{1, 1.5, \, -1.84, \, -0.26, \, -0.79, \, 1.92, \, 0.75, \, 0.17  \right\},
    \label{eq:theta_fid}
\end{split}
\end{equation}
where the fiducial noise parameters are set to their corresponding Poisson expectation (see Appendix \ref{sec:LO_noise_spectra}), inspired by galaxy number densities of current and upcoming spectroscopic surveys, and we choose to sample $10^{-3}P_\varepsilon$ instead of $P_\varepsilon$ in order to have all parameters to be of order unity, which is desirable for the neural density estimation. The fiducial Eulerian linear bias is set to $\bar{b}_1=1.5$, while $\bar{b}_{\text{tr}[M^{(1)}M^{(1)}]}$ and $\bar{b}_{\sigma \sigma}$ are determined from $\bar{b}_1$ through the bias relations (see Appendix \ref{sec:bias_relations}). 

Regarding $\bar{b}_{\nabla^2\delta}$, we expect its value to be close to zero, but with variance between two and three times the Lagrangian radius of an average Euclid galaxy. To set its fiducial value, we first find the minimum mass $M_{\text{min}}$ which satisfies the integral
\begin{equation}
    \left\langle b_O(z,M_{\text{min}}) \right\rangle = \int_{M_{\text{min}}}^{\infty} dM \frac{dn(z,M)}{dM} b_O(z,M)
    \label{eq:mean_bias}
\end{equation}
for $\left\langle b_1(z,M_{\text{min}}) \right\rangle = \bar{b}_1=1.5$ using the Tinker mass function for $n(z,M)$ \cite{Tinker_2008} and the Tinker linear bias predictions for $b_1(z,M)$ \cite{Tinker_2010}. We then use Eq. \eqref{eq:mean_bias} for the higher-derivative bias evaluated at the obtained $M_{\text{min}}$, where $b_{\nabla^2\delta}(z,M)$ is determined by $-R_L^2(M)/2.5$ \cite{Lazeyras_2019}, to set its fiducial value as the calculated mean, i.e., $\bar{b}_{\nabla^2\delta} = \left\langle b_{\nabla^2\delta}(z,M_{\text{min}}) \right\rangle$. 

We compare our results to a Fisher forecast, which approximates the posterior with a Gaussian via the ensemble average of its curvature of around the maximum. The Fisher information matrix for the likelihood is defined as
\begin{equation}
    {F}_{\alpha\beta}^{(\mathcal{L})} \equiv - \left\langle \frac{\partial^2 \ln \mathcal{L}}{\partial\theta_\alpha\partial\theta_\beta}\right\rangle\bigg|_{\bar{\boldsymbol{\theta}}},
\end{equation}
which for a Gaussian likelihood reduces to 
\begin{equation}
    {F}_{\alpha\beta}^{(\mathcal{L})} = \sum_{i,j=1}^D \left[ \frac{\partial x_i}{ \raisebox{-2pt}{ $\partial \theta_\alpha $} }
    \left(\text{Cov}[\textbf{x}]\right)^{-1}_{ij}
    \frac{\partial x_j}{\raisebox{-2pt}{ $\partial \theta_\beta$}}\right]\Bigg|_{\bar{\boldsymbol{\theta}}} ,
\end{equation}
where the mean of the derivatives as well as the covariance of the data vector ${\textbf{x}}$ over initial conditions realizations are evaluated at the fiducial parameters values $\bar{\boldsymbol{\theta}}$. We also include the prior Fisher matrix $F^{(p)}_{\alpha\beta}=\delta_{\alpha\beta}/\sigma_\alpha^2$, where $\delta_{\alpha\beta}$ is the Kronecker delta function and $\sigma_\alpha$ is the standard deviation considered for the parameter $\theta_\alpha$ in a Gaussian prior. The total Fisher information matrix for the posterior is then $F=F^{(\mathcal{L})}+F^{(p)}$. The marginal posterior in any subspace of the parameter space is then controlled by the corresponding restriction of $F^{-1}$. Specifically, the 1-sigma uncertainty on $\theta_\alpha$ is approximated by $\sqrt{(F^{-1})_{\alpha\alpha}}$. As previously discussed, the power spectrum and bispectrum are not exactly Gaussian distributed, and one of the main goals of this paper is to explore the final parameter posterior when this assumption is relaxed. However, the Fisher prediction provides a guideline for interpreting our results.

Since we aim to use NPE instead of sequential methods for amortabilizity and since SBI requires many simulations around high density posterior regions, a careful choice of priors (i.e. neither too wide nor too narrow) is crucial. We hence choose the mean and variance of our priors in the following way. First, we run SNPE with batches of $10^3$ simulations starting from the prior
\begin{equation}
    \begin{split}
    &\alpha\sim\mathcal{N}(1.2,1^2), \quad b_1 \sim \mathcal{N}(1.25, 2^2), \quad 
    b_{\sigma\sigma} \sim \mathcal{N}(0, 2^2),  \quad
    b_{\text{tr}[M^{(1)}M^{(1)}]} \sim \mathcal{N}(0, 2^2), \\
    &b_{\nabla^2\delta} \sim \mathcal{N}(0, 25^2), \quad 10^{-3} \,P_{\varepsilon} \sim \mathcal{N}(2, 1^2), \quad 
    c_{\varepsilon\delta} \sim \mathcal{N}(0, 2^2),  \quad
    c_{\varepsilon^2} \sim \mathcal{N}(0, 2^2),
    \end{split}
    \label{eq:snpe_prior}
\end{equation}
where we choose the centers of $\alpha$ and $b_1$ in such a way that their multiplication equals $1.5 = \bar{\alpha} \times \bar{b}_1$. After SNPE convergence, we sample from the final posterior to choose the prior for NPE to be a Gaussian centered on the sample mean with variance given by at least two times the SNPE sample variance. For all inference cases, we overplot the prior with the posterior to guarantee that the latter is not bounded by the prior ranges. We refer the reader to Appendix \ref{sec:sec_bias_tests} for examples comparing SNPE and NPE.

\subsubsection{Fixed cosmology: impact of the likelihood form}
\label{sec:alpha_fixed}

We start with the case where $\alpha$ is fixed to unity, while all the first and second order bias and noise parameters are sampled. We name as \textit{full SBI} results from the forward model where the measured data vector from \texttt{LEFTfield} are used directly for posterior estimation. We compare these results to a \textit{Gaussian-likelihood} model, where we generate samples from the data vector as $\textbf{x} \sim\mathcal{N}\left(\langle \textbf{x} \rangle, \text{Cov}[\bar{\textbf{x}}]\right)$, where $\langle \textbf{x} \rangle$ denotes the mean of the data vector over initial conditions realizations calculated at the proposed parameter $\boldsymbol{\theta}$ and the covariance is calculated for the data vector at the fiducial parameters $\bar{\textbf{x}}$. While the name Gaussian likelihood refers to the fact that the data is Gaussian distributed by construction, we emphasize that the posterior is also obtained with NPE as in the full-SBI case. We of course expect to obtain the same result when performing a Monte Carlo sampling based on the same Gaussian likelihood for the data vector. 

\begin{figure}
    \centering
    \includegraphics[width=0.49\textwidth]{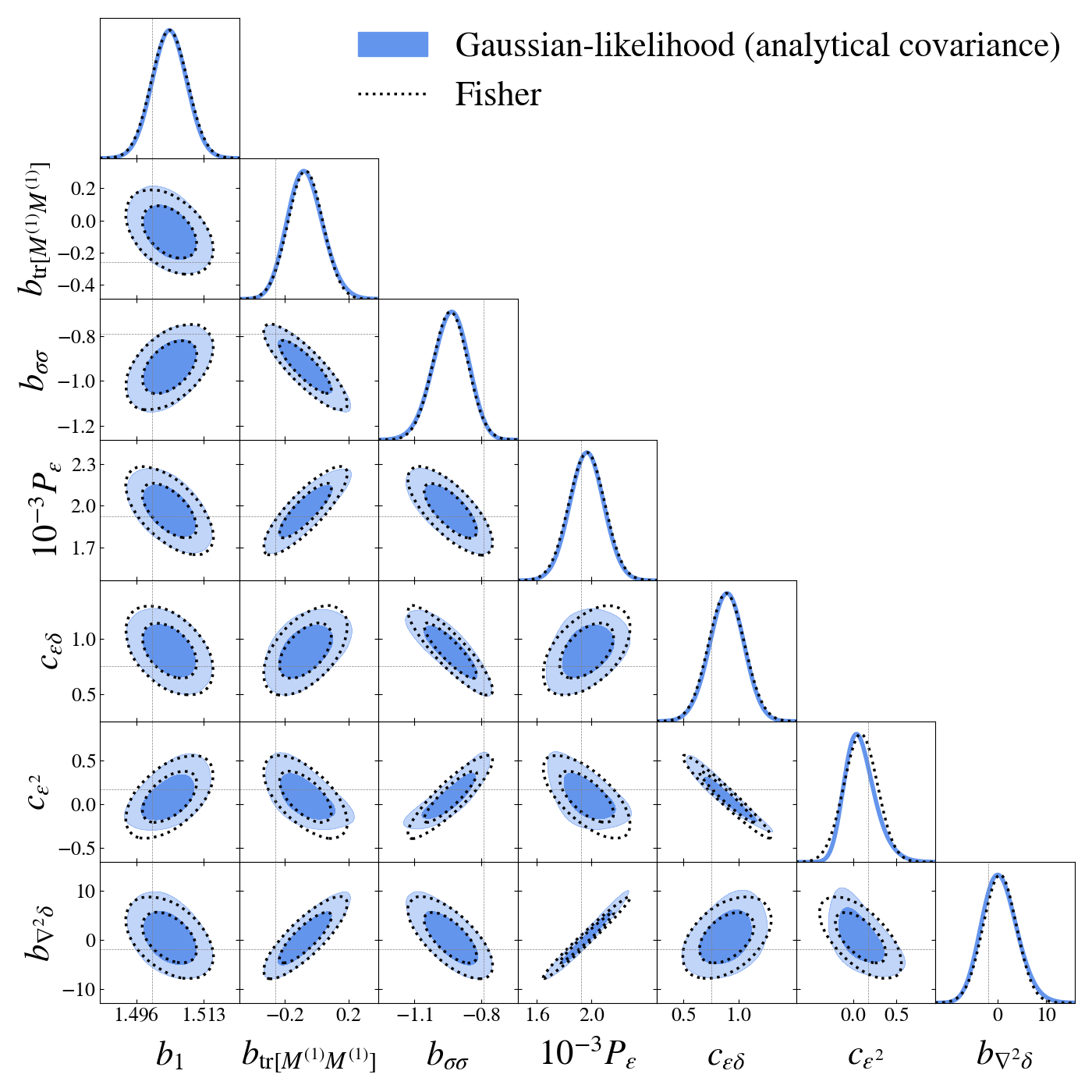}
    \includegraphics[width=0.49\textwidth]{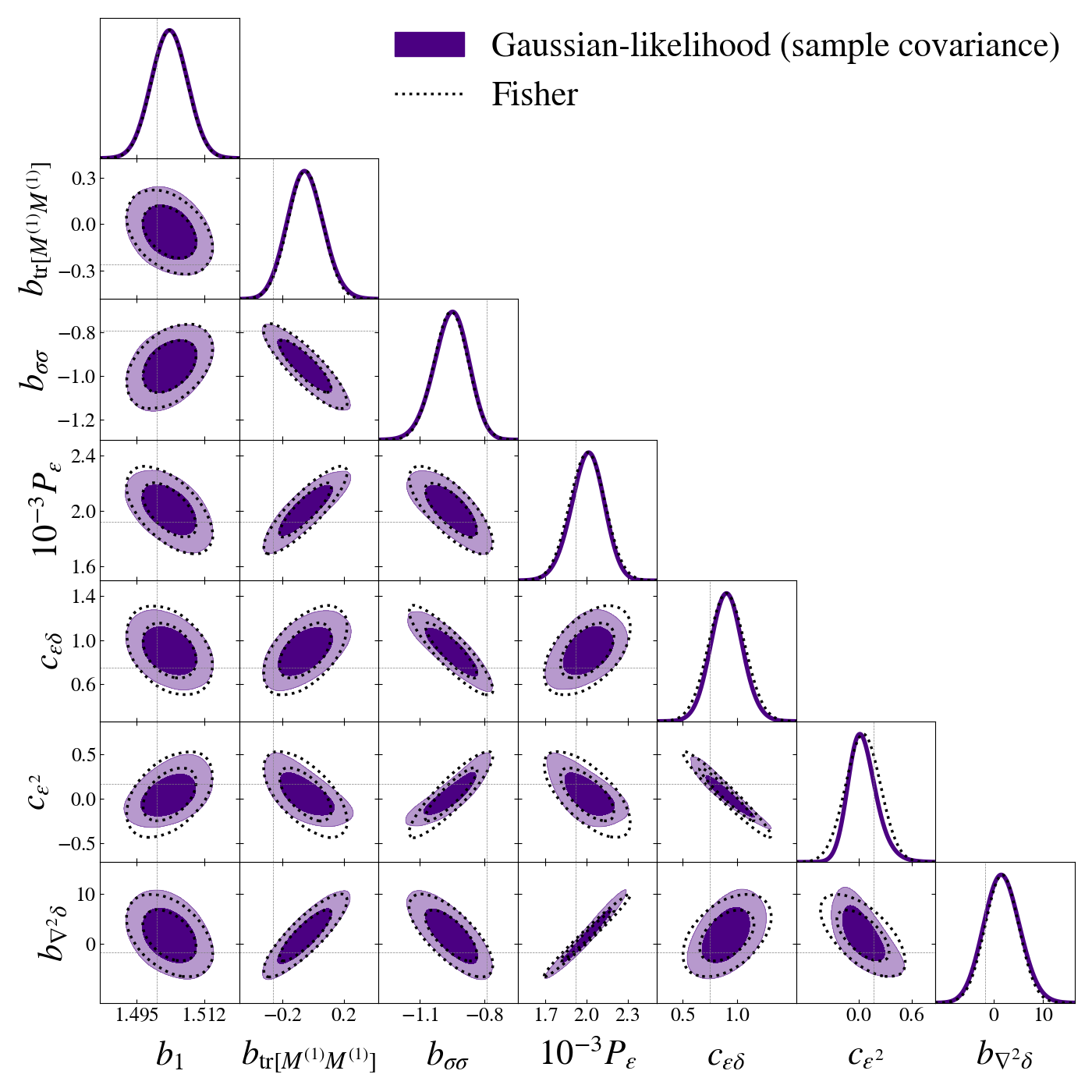}
    \caption{Parameter posterior for the case where the cosmological parameter $\alpha$ is fixed. The left and right contour plots correspond to the models where the data vector is sampled from a Gaussian with analytical and sample covariance, respectively. For all cases, the neural density estimator method NPE was used with a simulation budget of $N_\text{sim}=10^5$.}
    \label{fig:alphafix_gaussian_fisher}
\end{figure}

Regarding the covariance $\text{Cov}[\bar{\textbf{x}}]$, we consider two distinct cases. First, we consider an \textit{analytical}, diagonal covariance for the data vector, namely $\text{Cov}[\bar{P}(k)]=2\langle\bar{P}(k)\rangle^2/N_k$, where $\langle\bar{P}(k)\rangle$ is the mean of the fiducial power spectrum over initial conditions evaluated at fiducial parameters $\bar{\boldsymbol{\theta}}$ and $N_k$ is the total number of modes which lie in the respective $k$-bin, and $\text{Cov}[\bar{B}(k_1,k_2,k_3)]=s_B \, \langle\bar{P}(k_1)\rangle\langle\bar{P}(k_2)\rangle\langle\bar{P}(k_3)\rangle/k_f^3N_t$, where $s_B$ is the triangle shape symmetry factor (6, 2 and 1 for equilateral, isosceles and scalene triangles, respectively), $k_f=2\pi L^{-1}$ is the fundamental frequency and $N_t$ is the number of triangle configurations inside each triangle bin. Second, we measure the \textit{sample} covariance from $N_{\text{cov}}$ simulations with parameters fixed at their fiducial values.  

To summarize, in both \emph{Gaussian-likelihood} cases, the likelihood of the data vector is characterized completely by its first and second  moments (namely the mean and covariance), where the covariance is fixed to the fiducial point in parameter space. Compared to the analytical covariance, the sample covariance case adds more realism, since cross-correlations between different elements of the data vector (i.e. off-diagonal elements of the covariance) are included. In contrast, the \emph{full SBI} forward model captures the full distribution of the data vector, including its higher-order moments, and as function of the position in parameter space.

\begin{figure}
    \centering
    \includegraphics[width=0.8\textwidth]{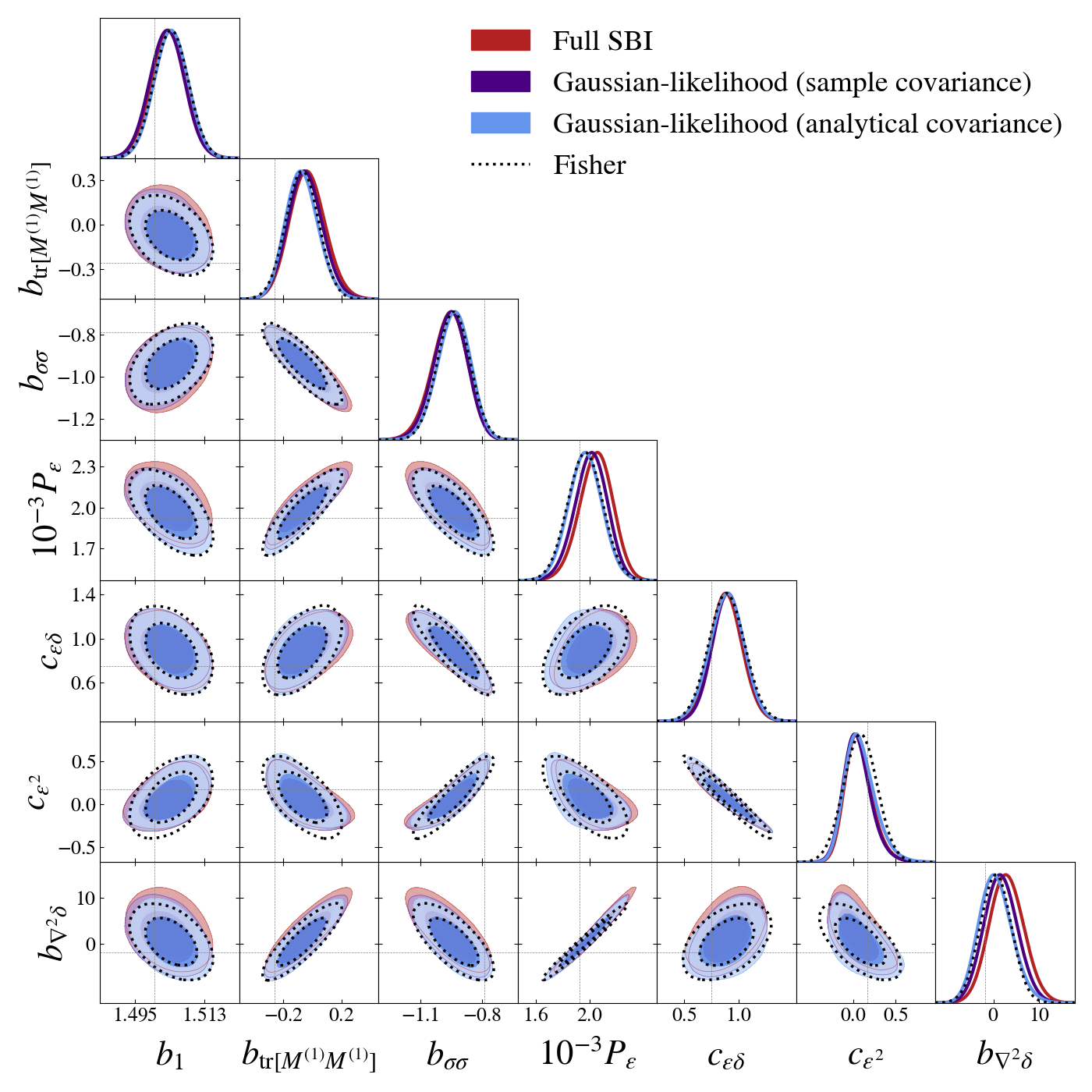}
    \caption{Parameter posterior for the case where the cosmological parameter $\alpha$ is fixed. Blue and purple contours correspond to the models where the data vector is sampled from a Gaussian likelihood with analytical and sample covariance, respectively, where $N_{\text{cov}}=10^4$. The red contours show the full SBI results, with no Gaussian assumption. Dotted lines indicate the Fisher prediction with sample covariance for reference. For all cases, the method NPE was used from a simulation budget of $N_\text{sim}=10^5$, scale cut of $k_{\text{max}}=\Lambda=0.1h$Mpc$^{-1}$ and data vector dimension $D=33$.}
    \label{fig:alphafix_models}
\end{figure}

We show in Figure \ref{fig:alphafix_gaussian_fisher} the estimated posteriors for the Gaussian likelihood model at $k_{\text{max}}=\Lambda=0.1h\text{Mpc}^{-1}$, which closely resembles the Fisher prediction. For all parameter posteriors considered in this work, we always sample $10^5$ posterior samples for plotting. The Fisher derivatives are taken numerically for the full case, with steps given by the standard deviation of the NPE posterior, while for the Gaussian-likelihood case it is evaluated analytically at the fiducial values from the mean cross-spectra of the bias expansion basis. The small deviations come from the fact that, although the likelihood and the prior are Gaussian, the posterior is not necessarily Gaussian, as the parameters enter nonlinearly in the prediction for the expectation value of the data vector. Note that, although the Fisher prediction is evaluated at the fiducial parameters points, we shift the mean of the Fisher contours to the maximum a posteriori posterior values for better comparison. This results shows that NPE can successfully recover the expected posterior in this idealized case, showing that this method is therefore robust for the data vector size and number of simulations used.

\begin{figure}
    \centering
    \includegraphics[width=0.8\textwidth]{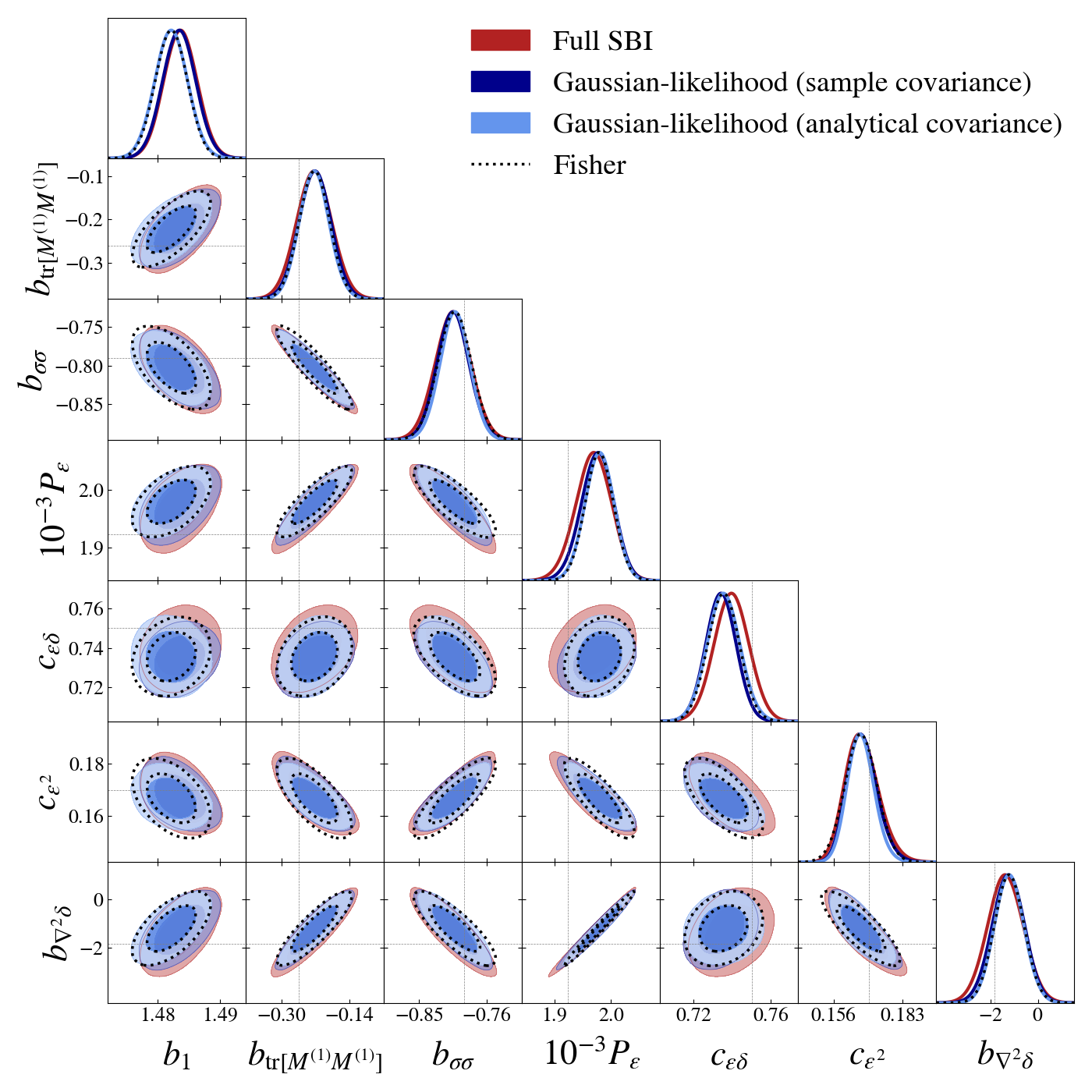}
    \caption{Same as Fig. \ref{fig:alphafix_models}, but for $k_{\text{max}}=\Lambda=0.2h$Mpc$^{-1}$ and $D=49$.}
    \label{fig:alphafix_models_L02}
\end{figure}

Figure \ref{fig:alphafix_models} compares the Gaussian-likelihood posteriors with the full-SBI case. First of all, we can notice that the Gaussian likelihood cases are very similar, which indicates that off-diagonal terms do not play an important role on the scales $k\leq\Lambda= 0.1 \, h \text{Mpc}^{-1}$ considered here. The difference between these and the full-SBI posteriors is slightly larger, but still small, which also indicates that the non-Gaussianity of the data vector also does not affect the constraints. Figure~\ref{fig:alphafix_models_L02} shows the corresponding result when going to smaller scales, $k\leq\Lambda= 0.2 \, h \text{Mpc}^{-1}$. The posterior differences are somewhat larger than on smaller scales, as expected, though still not dramatic.

\begin{figure}
    \centering
    \includegraphics[width=0.9\textwidth]{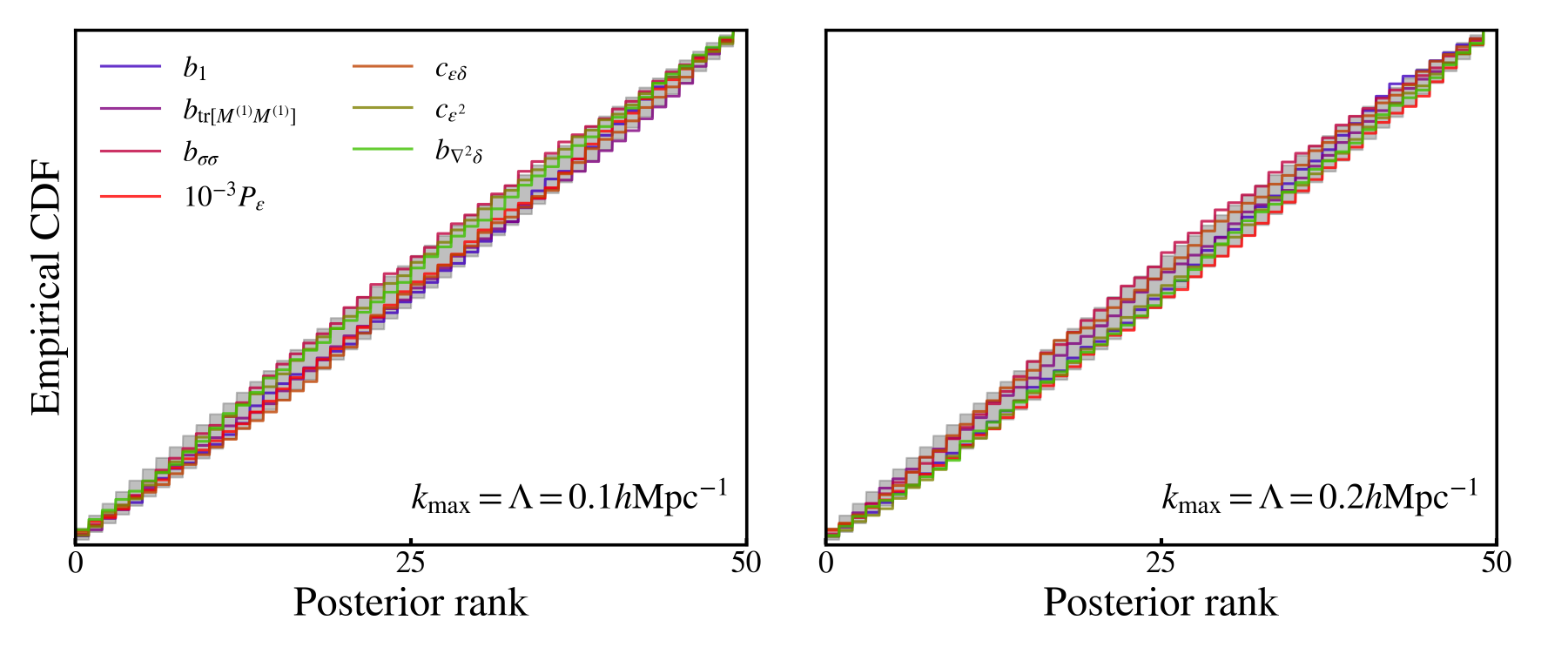}
    \includegraphics[width=\textwidth]{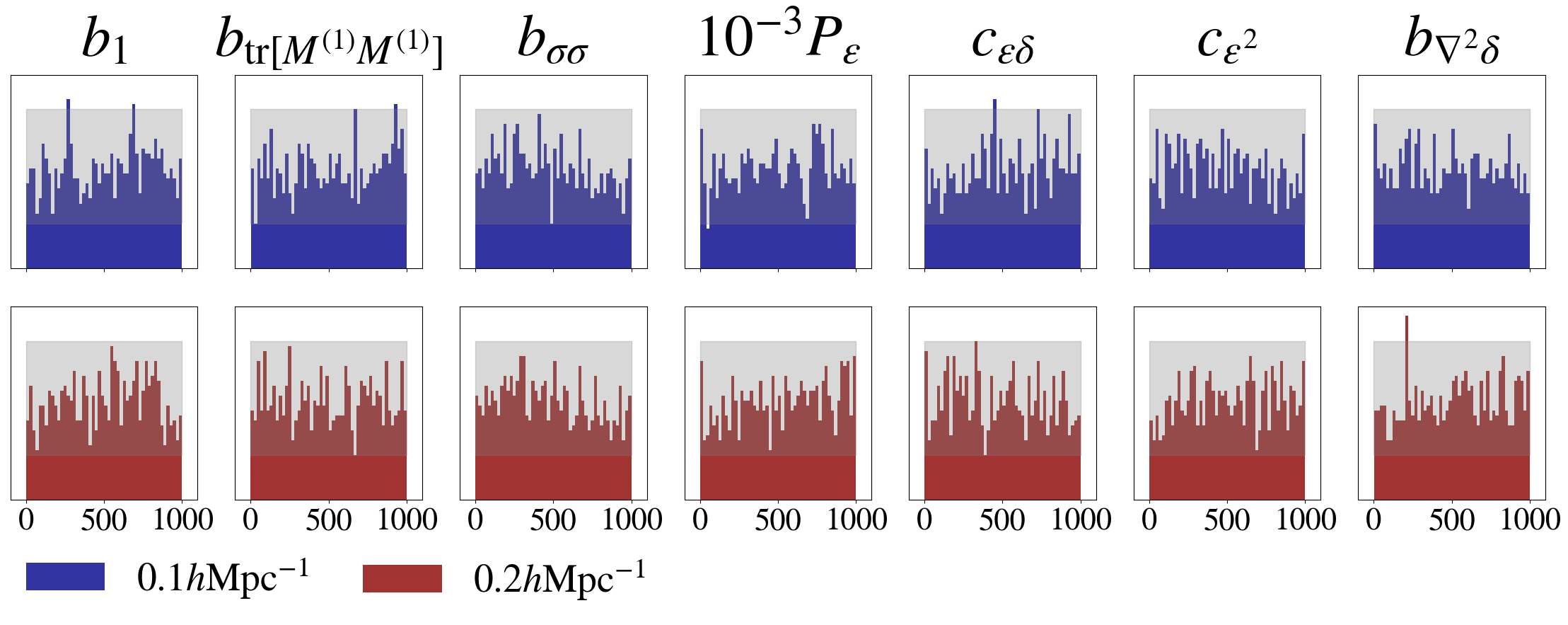}
    \caption{SBC (simulation-based calibration) tests for the full-SBI case where the cosmological parameter $\alpha$ is fixed with $N_\text{sim}=10^5$. The upper panels show the CDF of the ranks distributions for each parameter, where the grey area show the 95\% confidence interval of a uniform distribution, while the lower panels show the rank distribution, where the grey areas denote the 99\% confidence interval of a uniform distribution.}
    \label{fig:alphafix_coverage}
\end{figure}

\begin{figure}[h!]
    \centering
    \includegraphics[width=0.8\textwidth]{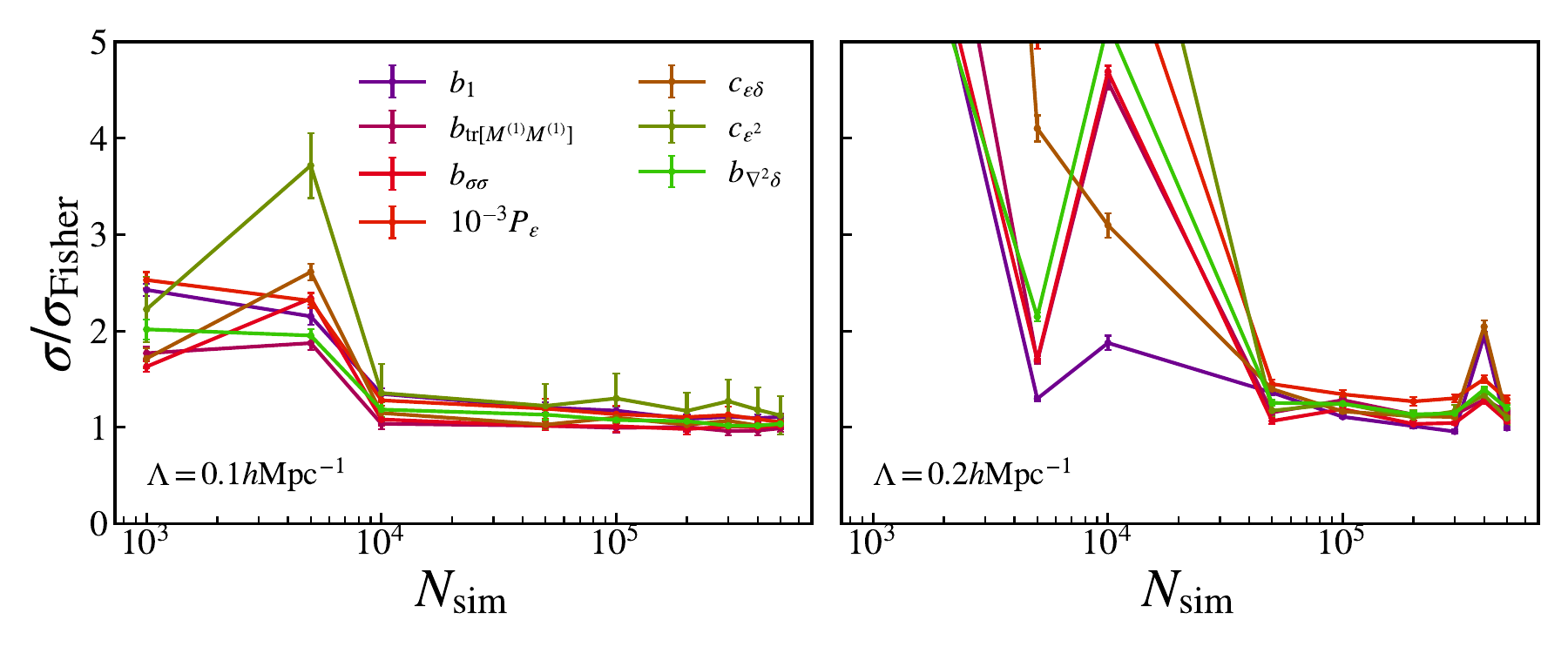}    
    \caption{Convergence of the full-SBI parameter posterior with the number of simulations $N_\text{sim}$ for the case where the cosmological parameter $\alpha$ is fixed. The standard deviation of the posterior samples are normalized by their respective Fisher prediction for better comparison. Note that it should not necessarily converge to one. The errors indicate posterior evaluation at 10 different data observations, with same fiducial parameters but different initial conditions realizations.}
    \label{fig:alphafix_convergence}
\end{figure}

We show the simulation-based calibration (SBC) tests of the posterior in Figure \ref{fig:alphafix_coverage} (see Appendix \ref{sec:sbc}). As previously discussed, a healthy posterior should lead to uniformly distributed rank statistics for all parameters, and we show two different visualizations which can help us in identifying possible problems with the estimated posterior. First, if the SBI underestimates the true posterior variance for some parameter, one expects a ``U-shaped'' rank distribution, or equivalently a CDF lying below the grey shaded area of the 95\% confidence interval of a uniform distribution. Conversely, if the posterior overestimates the errors, one would get a centrally peaked distribution, or a CDF above the grey area. Since our ranks are uniformly distributed and all CDFs lie inside the grey region, we conclude that our posterior passed the calibration test, although we re-emphasize that this is only a necessary and not a sufficient condition.

We show convergence of the obtained posterior with respect to the number of simulations for the full case in Figure \ref{fig:alphafix_convergence}. The results indicate that adding more simulations than $N_{\text{sim}}=10^5$, the simulation budget used in the previous figures, does not change the full-SBI posteriors. The conclusions hold equivalently for the Gaussian likelihood cases. Note that the sudden upturns for a low simulation budget only indicates that the budget was not sufficient for convergence, and that running SBI again on the same dataset could lead to different posteriors. 

\clearpage
\subsubsection{Inferring cosmology}
\label{sec:alpha_free}

\begin{figure}[t]
    \centering
    \includegraphics[width=0.9\textwidth]{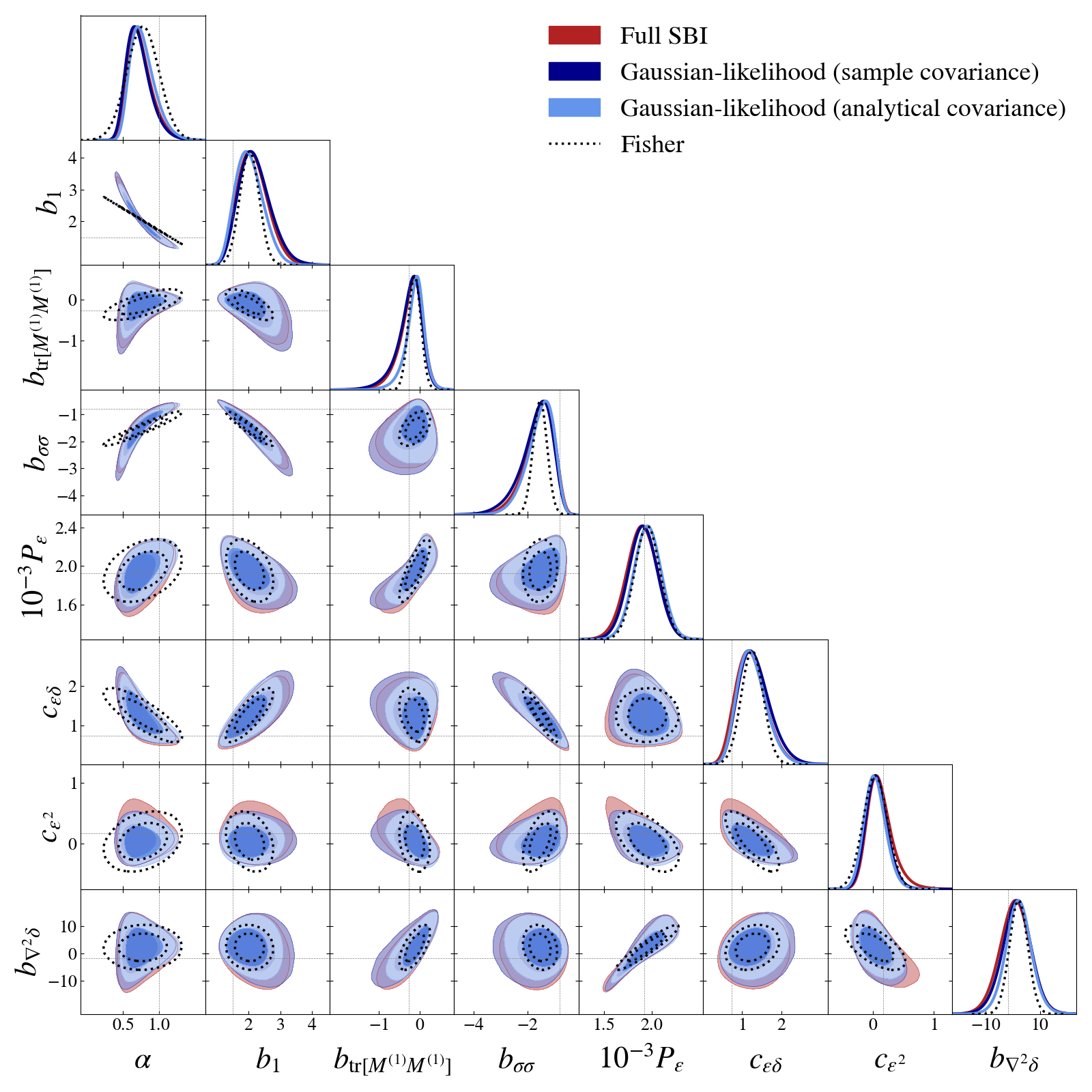}
    \caption{Parameter posterior for the case where the cosmological parameter $\alpha$ is sampled together with the bias and noise parameters. The contour colors, simulation budget and data vector size are the same as in Fig. \ref{fig:alphafix_models}. For all cases, the method NPE was used from a simulation budget of $N_\text{sim}=10^5$, scale cut of $k_{\text{max}}=\Lambda=0.1h$Mpc$^{-1}$ and data vector dimension $D=33$.}
    \label{fig:alphafree_contours_l01}
\end{figure}

We now turn to constraining the bias and noise parameters together with the cosmological parameter $\alpha$, where we recall that $\alpha\equiv\sigma_8/\sigma_8^{\text{fid}}$ and set its fiducial value to $\bar{\alpha}=1$. We show in Figure \ref{fig:alphafree_contours_l01} how we are able to infer these parameters independently with the help of the bispectrum at $k_{\text{max}}=0.1h$Mpc$^{-1}$, and that SBI is successful in this context even considering the dimensionality of the problem and the nontrivial parameter degeneracies.

\begin{figure}[t]
    \centering
    \includegraphics[width=0.9\textwidth]{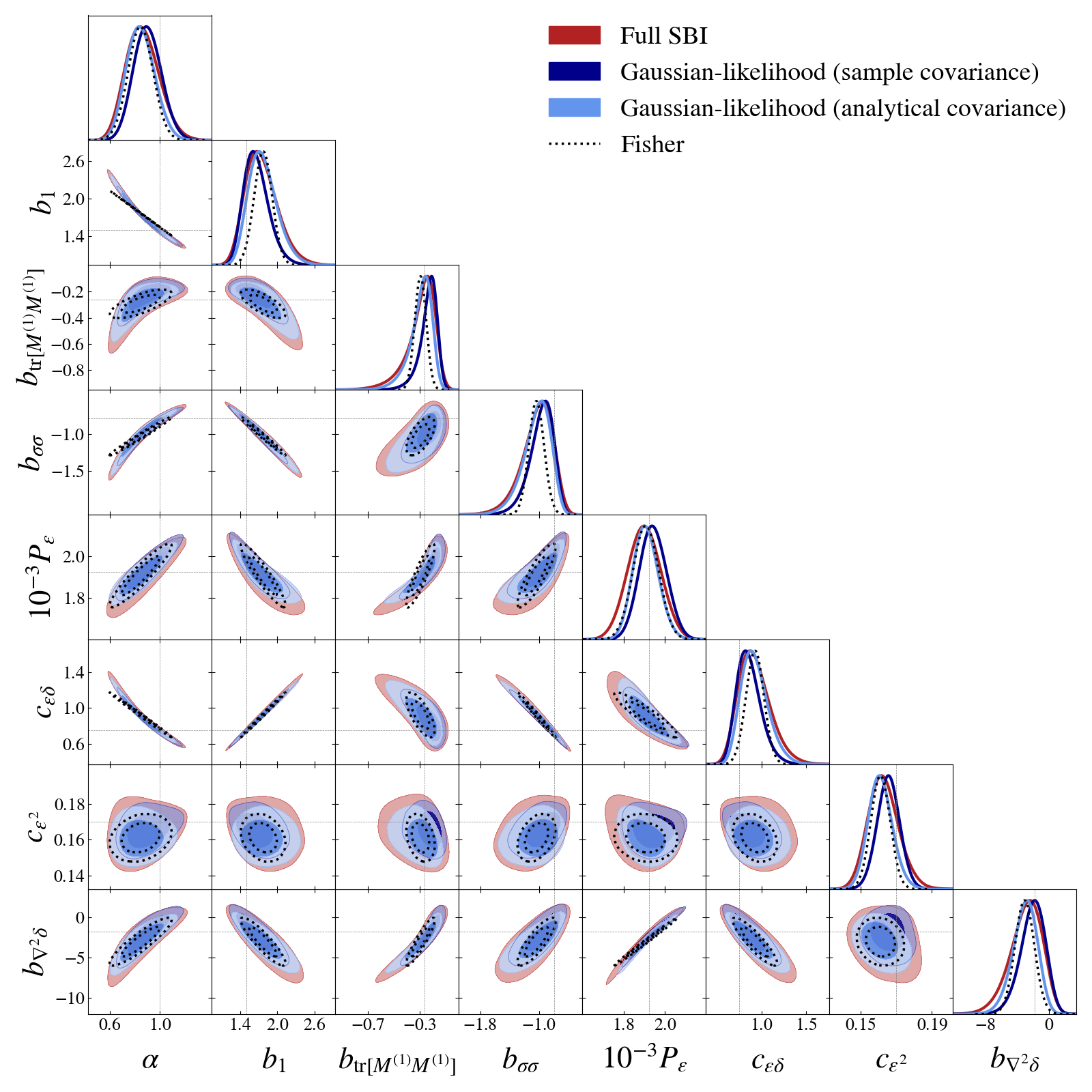}
    \caption{Same as Fig. \ref{fig:alphafree_contours_l01}, but for $k_{\text{max}}=\Lambda=0.2h$Mpc$^{-1}$ and $D=49$.}
    \label{fig:alphafree_contours_l02}
\end{figure}

We can notice that the analytical covariance underestimates the errors in this case, a result that has already been discussed in the galaxy clustering literature \cite{2019MNRAS.482.4883C,2019MNRAS.485.2806B}. We however find no significant difference between the Gaussian likelihood case with sample covariance and the full-SBI one, a result which is in agreement with the current intuition that adding off-diagonal terms to the covariance would be more important than considering a non-Gaussian likelihood. Our conclusion is that, for this particular case, the non-Gaussianity of the data vector (especially at low-$k$ modes) does not lead to any considerable effect on the posterior densities. The conclusion still holds for $k_{\text{max}}=0.2h$Mpc$^{-1}$, as displayed in Figure \ref{fig:alphafree_contours_l02}, although differences between full-SBI and Gaussian-likelihood results become more noticeable.
The 1-$\sigma$ errobars for the full-SBI, Gaussian-likelihood with sample covariance, Gaussian-likelihood with analytical covariance and the Fisher forecast are 0.173, 0.173, 0.169 and 0.21 $k_{\text{max}}=0.1h$Mpc$^{-1}$, respectively, while 0.131, 0.111, 0.111 and 0.100 for $k_{\text{max}}=0.2h$Mpc$^{-1}$. 

\begin{figure}
    \centering
    \includegraphics[width=0.9\textwidth]{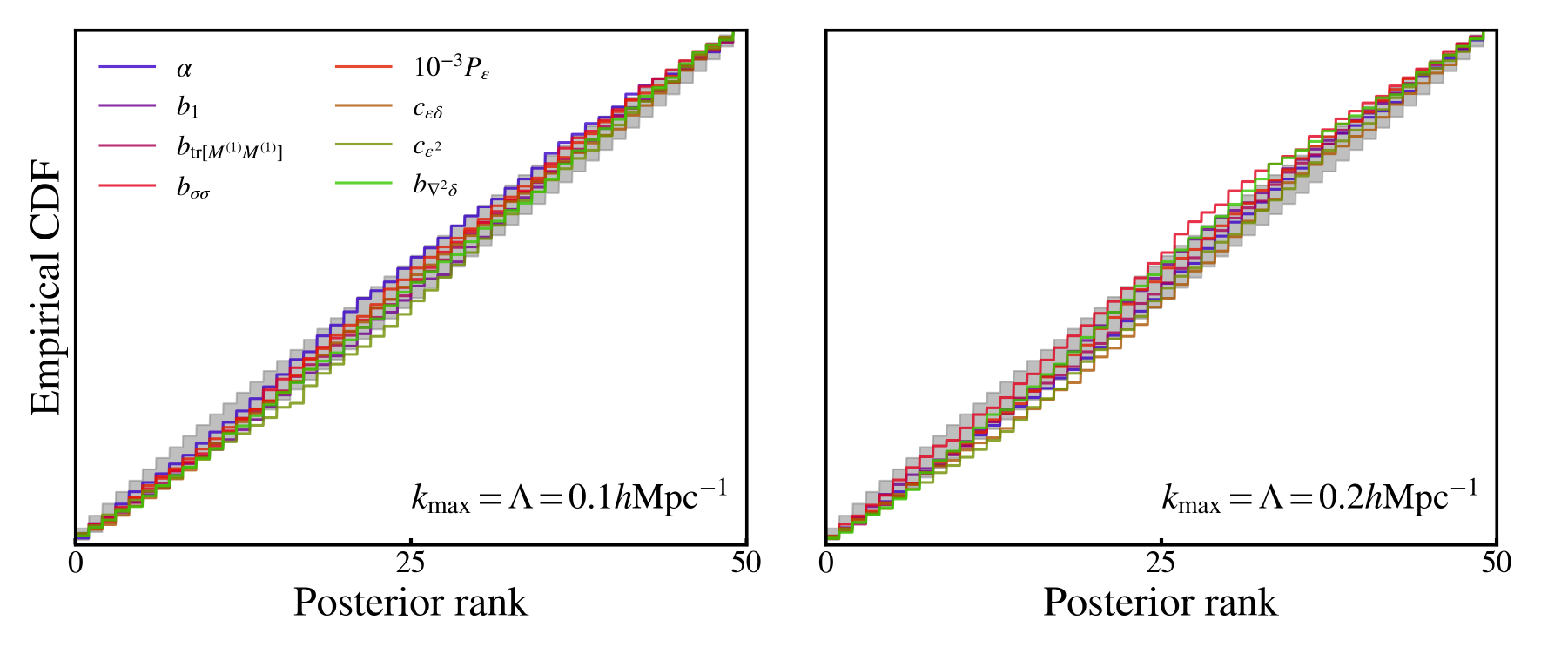}
    \includegraphics[width=1.0\textwidth]{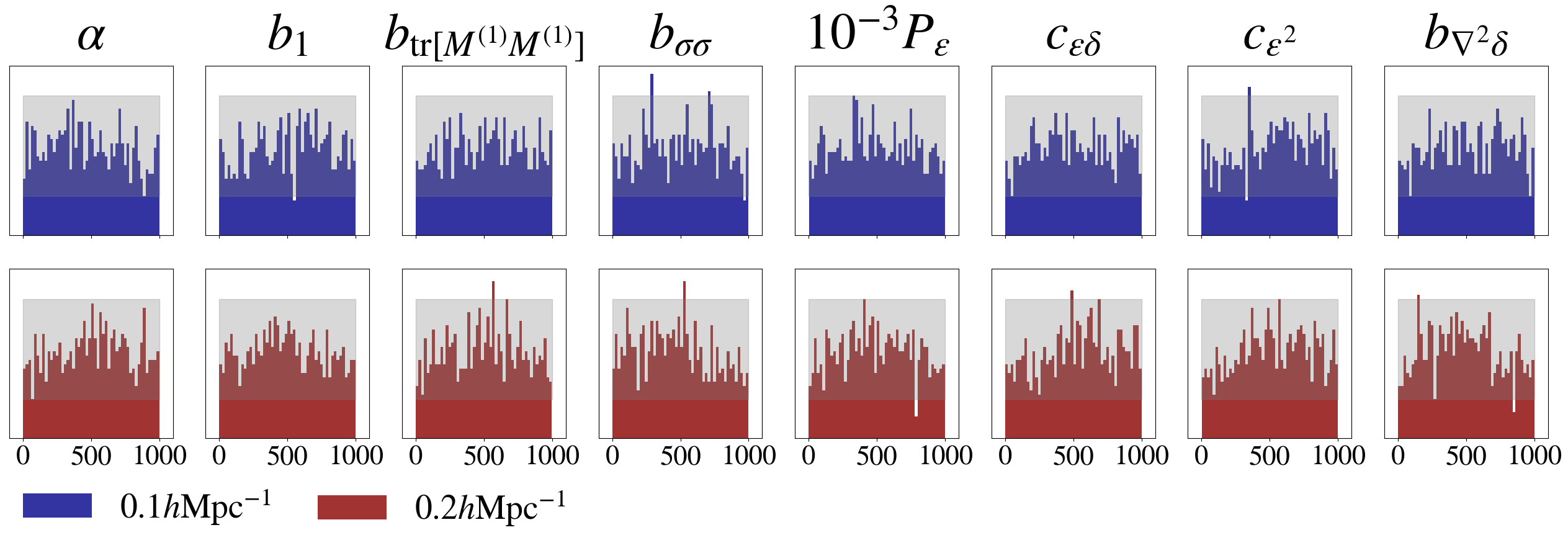}
    \caption{Same as Fig.~\ref{fig:alphafix_coverage} but for the case where the cosmological parameter $\alpha$ is sampled as well, with $N_\text{sim}=10^5$.}
    \label{fig:alphafree_coverage}
\end{figure}

\begin{figure}
    \centering
    \includegraphics[width=0.8\textwidth]{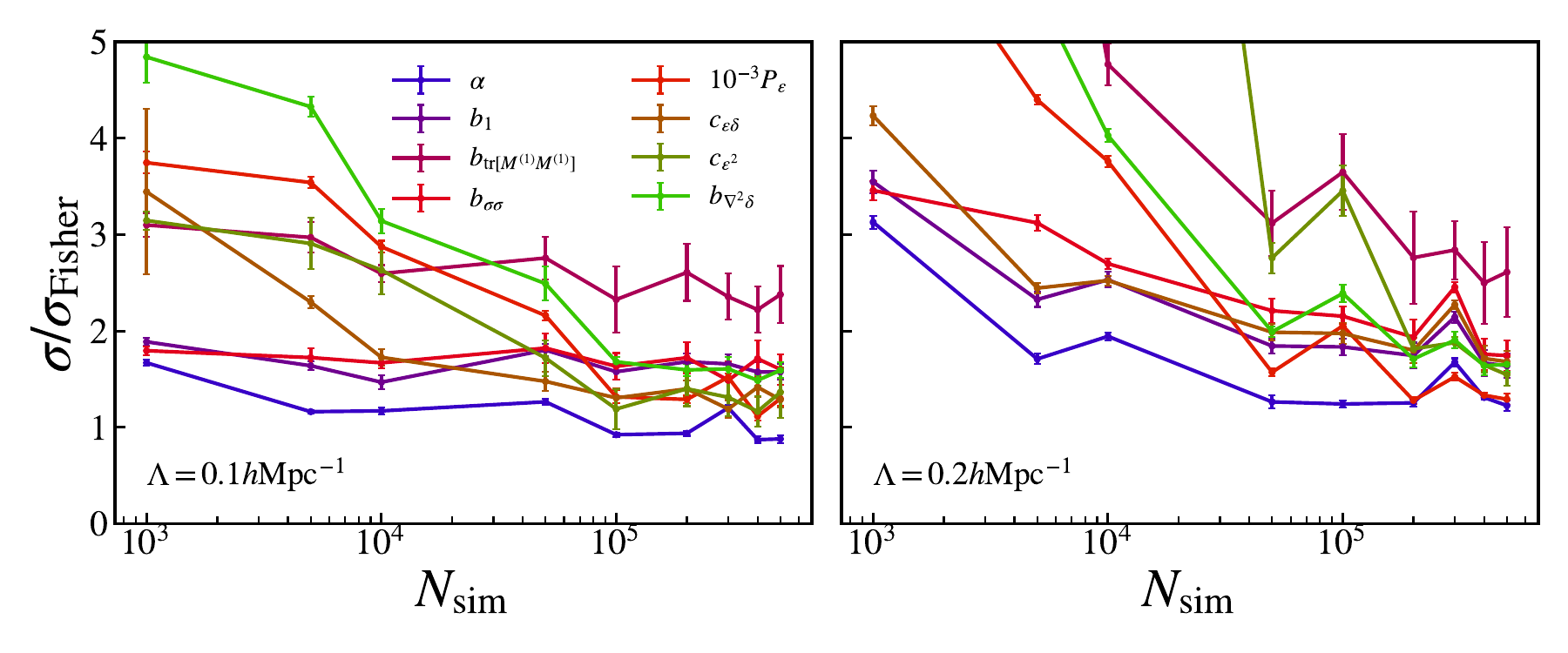}
    \caption{Same as Figure \ref{fig:alphafix_convergence}, but showing convergence for the full case where the cosmological parameter $\alpha$ is sampled.}
    \label{fig:alphafree_convergence}
\end{figure}

We emphasize here that this conclusion is highly problem-dependent, and it does not necessarily extend to other cosmological parameters, scale ranges or summary statistics. For example, since most of the constraining power on $f_{\text{NL}}$ comes from low-$k$ modes, one could expect it to be more sensitive to the Gaussian likelihood assumption. At the other extreme, on highly nonlinear small scales, one can also expect deviations from Gaussianity, while some summary statistics might not even be well described by an analytical form in any scale range.

In general, relaxing the Gaussian assumption in the likelihood is a non-trivial analytical task, but SBI methods provide us a straightforward route. The investigation of whether relaxing the Gaussian likelihood assumption leads to an increase of the error bars has been previously discussed in the literature; for example, \cite{2019MNRAS.485.2956H} reported no significant difference when using the power spectrum only. In contrast, Figure 3 of \cite{Leclercq_2021} show that when using the two-point correlation function, the posterior obtained by SBI displays larger error bars than when considering a Gaussian likelihood in the context of lognormal fields.

The SBC results are shown in Figure \ref{fig:alphafree_coverage}, which indicate that our posterior estimates pass the calibration test also for this case. 
The convergence is analyzed in Figure \ref{fig:alphafree_convergence}. We can notice a slower convergence with respect to the number of simulations when compared to the case where $\alpha$ is fixed, which might be related to the fact that the two-dimensional contours of $\alpha$ and the other parameters are more non-Gaussian in this case, due to the increased degeneracies relative to the $\alpha$-fixed case, and therefore the normalizing flow needs more simulations to fit the posterior.

We show in Appendix \ref{sec:sec_bias_tests} an analysis of how our results can be affected by data vector normalization and the number of sample covariance estimates for the Gaussian-likelihood case, besides providing a comparison of bispectrum with power-spectrum-only constraints. We also compare the results with SBI density estimators other than NPE (namely SNPE and NLE), as well as the impact of a more complex network architeture or using an ensemble of networks instead of a single realization for posterior estimation.

\section{Conclusions}
\label{sec:conclusion}

We have explored for the first time how SBI performs within the context of a forward model based on the EFTofLSS and the bias expansion with the goal of constraining cosmological, noise and bias parameters from the galaxy power spectrum and bispectrum. First, we have demonstrated that SBI can successfully recover the expected posterior in controlled cases where the data vector is sampled from a Gaussian likelihood. Considering an Euclid-like mock tracer sample, we conclude that the non-Gaussianity of the lowest order $n$-point functions on large scales does not impact the constraints on $\sigma_8$ for the scale ranges considered in this work, namely $k_{\text{max}}=0.1h\text{Mpc}^{-1}$ and $0.2h\text{Mpc}^{-1}$. In these cases, adding off-diagonal terms to the covariance has a larger impact than relaxing the Gaussian likelihood assumption. Our estimated posteriors passed the SBC test and, within this specific context, we estimated the number of simulations needed for convergence to be of the order of $10^5$ simulations.

It is important to emphasize that, while SBI serves as a versatile and powerful tool for cosmological inference, our utilization of machine-learning is strictly limited to the statistical aspect of our inference process, specifically density estimation. This entails essentially a fitting of the posterior distribution or likelihood function derived from a set of simulations, allowing us to overcome the limitations of standard inference techniques that rely on analytical approximations for the selected summary statistics. Most importantly, this work is performed within the framework of a forward model rigorously constructed to ensure accuracy on large scales. In this context, our understanding of the underlying physics benefits from years of dedicated study within the EFTofLSS community, providing us with a high level of control and intuitive comprehension, especially when employing the power spectrum and bispectrum as summary statistics. 

It is paramount to validate and establish the reliability of SBI in well-understood regimes, such as those under consideration in this study. This validation process represents a crucial first step before delving into more complicated scenarios, where intuitive comprehension may be lacking. In addition, knowledge of an explicit likelihood that is at least valid for some elements of the data vector (for example on very large scales) allows for the possibility for explicit combinations of analytical likelihoods and SBI \cite{Modi_2023a}.

As the next direct step of this work, we will test our inference pipeline on dark-matter halos from N-body simulations and test different compression schemes other than $n$-point functions. It would also be interesting to investigate the impact of the details of the forward model that differ in our approach and current EFT-based analysis techniques based from codes such as CLASS-PT or PyBird (see the discussion in \cite{Rubira:2023vzw}). We also aim to use SBI to investigate whether the low-$k$ non-Gaussianities of the lowest order $n$-point functions impact the inference of cosmological parameters related to primordial non-Gaussianities. 

In future work, we plan to extend our analysis to redshift space, using the \texttt{LEFTfield} forward model extension presented in \cite{Stadler_2023}, and to sample more cosmological parameters and to improve the realism of the forward model by incorporating systematic effects and masks in \texttt{LEFTfield}. We are particularly interested in comparing the results from field-level cosmological inference, which explicitly marginalizes over the initial conditions using HMC, to SBI inference, both using $n$-point functions as summary statistics and other ``field-level'' compression mechanisms which learn the summary statistics directly from the galaxy density field.

Regarding our SBI pipeline, we are planning to explore prior truncation schemes \cite{Miller_2021, Karchev_2023}, using techniques such as the Sobol' sequence \cite{Sobol_97} and active learning \cite{Alsing_2019a}. We would also like to improve our density estimation methods by hyperparameter tuning, besides trying other SBI algorithms such as Neural Ratio Estimation \cite{Hermans_2019, Durkan_2020, Miller_2021, Miller_2022, Delaunoy_2022}. In particular, it would be interesting to test Truncated Marginal Ratio Estimation (TMNRE) \cite{Miller_2021, Cole_2022} for marginalizing over the bias parameters. Lastly, it would be interesting to also explore other SBI diagnostics besides SBC \cite{Zhao_2021, Linhart_2022, Lemos_2023a}.

\acknowledgments

We would like to thank Ivana Babi\'c, Sam Goldstein, Eiichiro Komatsu, Andrija Kosti\'c, Nhat-Minh Nguyen, Henrique Rubira, Julia Stadler and Rodrigo Voivodic for valuable discussions during the preparation of this publication. Special thanks to Martin Reinecke for his insights and guidance regarding the optimization of our bispectrum estimator implementation.

\newpage 

\appendix

\section{Leading order stochastic bispectrum}
\label{sec:LO_noise_spectra}

The leading order stochastic contribution for the galaxy bispectrum in \texttt{LEFTfield} can be written as
\begin{equation}
    \begin{split}
    \big\langle\delta_g(\boldsymbol{k}_1)\delta_g(\boldsymbol{k}_2)\delta_g(\boldsymbol{k}_3)\big\rangle \Big|_{\text{stoch.}}^{\text{LO}} &= 
    c_{\varepsilon^2} \big\langle \varepsilon(\boldsymbol{k}_1)\varepsilon(\boldsymbol{k}_2)[\varepsilon^2](\boldsymbol{k}_3) + \text{2 perm.} \big\rangle \\
    &+ c_{\varepsilon\delta} \big\langle \delta_{g,\text{det}}(\boldsymbol{k}_1)\varepsilon(\boldsymbol{k}_2)[\varepsilon\delta](\boldsymbol{k}_3) + \text{5 perm.}\big\rangle,
    \end{split}
    \label{eq:LO_stoch_bispectrum}
\end{equation}
where the brackets denote multiplication in real space. Higher-order contributions are controlled by $c_{\varepsilon^2}c_{\varepsilon\delta}$ or more than one power of $c_{\varepsilon^2}$ or $c_{\varepsilon\delta}$, which are expected to be small (see below). Note that terms with only one power of $\delta_{g,\text{det}}\sim\delta$, $\delta$, $\varepsilon$ or $\varepsilon^2$ vanish due to the zero mean of these operators. Now we calculate the first contribution of Eq. \eqref{eq:LO_stoch_bispectrum} as
\begin{equation}
    \begin{split}
    \big\langle \varepsilon(\boldsymbol{k}_1)\varepsilon(\boldsymbol{k}_2)[\varepsilon^2](\boldsymbol{k}_3) \big\rangle 
    &= \int_{\boldsymbol{q}} \big\langle \varepsilon(\boldsymbol{k}_1)\varepsilon(\boldsymbol{k}_2)\varepsilon(\boldsymbol{q})\varepsilon(\boldsymbol{k}_3-\boldsymbol{q}) \big\rangle \\
    &= \int_{\boldsymbol{q}} \Big[ \delta_D(\boldsymbol{k}_1+\boldsymbol{q}) P_\varepsilon(\boldsymbol{k}_1) \delta_D(\boldsymbol{k}_2+\boldsymbol{k}_3-\boldsymbol{q}) P_\varepsilon(\boldsymbol{k}_2) \\
    &\quad + \delta_D(\boldsymbol{k}_1+\boldsymbol{k}_3-\boldsymbol{q}) P_\varepsilon(\boldsymbol{k}_1) \delta_D(\boldsymbol{k}_2+\boldsymbol{q}) P_\varepsilon(\boldsymbol{k}_2) \Big] \\
    & = 2 \delta_D(\boldsymbol{k}_1+\boldsymbol{k}_2+\boldsymbol{k}_3) P_\varepsilon(\boldsymbol{k}_1) P_\varepsilon(\boldsymbol{k}_2) \\
    & = 2 \delta_D(\boldsymbol{k}_1+\boldsymbol{k}_2+\boldsymbol{k}_3) P_\varepsilon^2,
    \end{split}
\end{equation}
where $\int_{\boldsymbol{q}}\equiv\int d\boldsymbol{q}/(2\pi)^3$ and on the last line we assumed that $P_\varepsilon(k)=P_\varepsilon, \, \forall \, k$. The second contribution in turn reads
\begin{equation}
    \begin{split}
    \big\langle \delta_{g,\text{det}}(\boldsymbol{k}_1)\varepsilon(\boldsymbol{k}_2)[\varepsilon\delta](\boldsymbol{k}_3) \big\rangle &= 
    \int_{\boldsymbol{q}} \big\langle \delta_{g,\text{det}}(\boldsymbol{k}_1)\varepsilon(\boldsymbol{k}_2)\varepsilon(\boldsymbol{q})\delta(\boldsymbol{k}_3-\boldsymbol{q}) \big\rangle \\
    & = \int_{\boldsymbol{q}} \Big[ \delta_D(\boldsymbol{k}_1+\boldsymbol{k}_3-\boldsymbol{q}) P_{\delta_{g,\text{det}}, \delta}(\boldsymbol{k}_1) \, \delta_D(\boldsymbol{k}_2+\boldsymbol{q}) P_\varepsilon(\boldsymbol{k}_2) \Big] \\
    & = \delta_D(\boldsymbol{k}_1+\boldsymbol{k}_2+\boldsymbol{k}_3) P_{\delta_{g,\text{det}}, \delta}(\boldsymbol{k}_1) P_\varepsilon(\boldsymbol{k}_2) \\
    & = \delta_D(\boldsymbol{k}_1+\boldsymbol{k}_2+\boldsymbol{k}_3) b_1 P_\varepsilon P_{m}(\boldsymbol{k}_1),
    \end{split}
\end{equation}
where on the last line we additionally assumed that, at leading order, $\delta_{g,\text{det}}(\boldsymbol{k}) = b_1 \delta(\boldsymbol{k})$, and $P_m$ is the evolved matter power spectrum. Collecting all terms, Eq. \eqref{eq:LO_stoch_bispectrum} can be written as
\begin{equation}
    \big\langle\delta_g(\boldsymbol{k}_1)\delta_g(\boldsymbol{k}_2)\delta_g(\boldsymbol{k}_3)\big\rangle^{\prime} \Big|_{\text{stoch.}}^{\text{LO}} = 6\,c_{\varepsilon^2}P_\varepsilon^2 + 2\,b_1c_{\varepsilon\delta}P_{\varepsilon}\,\big(P_{m}(\boldsymbol{k}_1)+2\text{ }\text{perm.}\big),
\end{equation}
where the prime denotes the Dirac delta function. We can now do the matching between the above equation and Eq. \eqref{eq:LO_stoch_spectra} to obtain that
\begin{equation}
    6\,c_{\varepsilon^2}P_\varepsilon^2 = B_\varepsilon, \quad c_{\varepsilon\delta}P_{\varepsilon} = P_{\varepsilon\varepsilon_\delta}.
\end{equation}
In the Poisson limit, it is expected that $P_\varepsilon=n_g^{-1}$, $B_\varepsilon=n_g^{-2}$ and $P_{\varepsilon\varepsilon_\delta}=b_1 n_g^{-1}/2$ \cite{Schmidt_2016}, where $n_g$ is the galaxy comoving number density. We therefore arrive at the values
\begin{equation}
    c_{\varepsilon^2} \stackrel{\text{Poisson}}{=} \frac{1}{6}, \quad c_{\varepsilon\delta} \stackrel{\text{Poisson}}{=} \frac{b_1}{2},
\end{equation}
which will be used for our fiducial values, as in Eq. \eqref{eq:theta_fid}. Note that the leading order stochastic spectra are only used here in order to determine the fiducial value of the stochastic parameters, and the forward model takes into account all loop contributions for cosmological inference.

\section{Bias relations}
\label{sec:bias_relations}

The bias expansion is constructed upon a sum of operators at given order in perturbation theory at a certain given time, where one can always relate a given basis of operators one to another. From the fiducial value of the linear bias $b_1$, which is related to the Eulerian basis $\boldsymbol{O}_E=\{\delta, \delta^2, K^2\}$, where $K^2\equiv(K_{ij})^2$ and
\begin{equation}
    K_{ij}(\boldsymbol{k}) \equiv \left[ \frac{k_ik_j}{k^2} - \frac{1}{3}\delta_{ij} \right]\delta(\boldsymbol{k}),
\end{equation}
we wish to determine a physically-motivated fiducial value for the second-order bias parameters associated to the Lagrangian operators $\boldsymbol{O}_L=\{\text{tr}[\boldsymbol{M}^{(1)}], \text{tr}[\boldsymbol{M}^{(1)}]^2,\text{tr}[\boldsymbol{M}^{(1)}\boldsymbol{M}^{(1)}]\}$. At leading order, these operators can be related as
\begin{equation}
\begin{split}
\sigma^2\equiv\big(\text{tr}[\boldsymbol{M}^{(1)}]\big)^2 &= \delta^2,
\qquad
\text{tr}[\boldsymbol{M}^{(1)}\boldsymbol{M}^{(1)}] = K^2 + \frac{1}{3}\delta^2.
\end{split}
\end{equation}
and assuming a local Lagrangian bias model with vanishing Lagrangian tidal bias, we can use the co-evolution relations \cite{Desjacques_2018} to obtain that the conversion between both sets of bias parameters reads
\begin{equation}
\label{eq:x}
\begin{split}
b_{\sigma \sigma} &= b_2 - \frac{4}{21}(b_1-1),\\
\qquad
b_{\text{tr}\left[M^{(1)} M^{(1)} \right] } & = \frac{1}{3}\Big(b_2-\frac{4}{21}(b_1-1)\Big),
\end{split}
\end{equation}
where the bias terms with no subscripts are the Eulerian bias. By setting $b_1=1.5$, $b_2$ is determined by the fitting formula $b_2(b_1)$ presented in Ref. \cite{Lazeyras_2016} as $b_2=-0.69$. Our second order bias terms are then set to the fiducial values $\bar{b}_{\sigma \sigma}=-0.79$ and $\bar{b}_{\text{tr}\left[M^{(1)} M^{(1)} \right]} = -0.26$, as in Eq. \eqref{eq:theta_fid}.
Note that the fiducial linear bias is $\bar{b}_1=1.5$, since it is the Eulerian linear bias by construction.

\section{Normalizing flows}
\label{sec:normalizing_flows}

Statistical inference has gained power with the advent of normalizing flows, a class of generative models that allows one to model complex probabilistic distributions. The key idea behind this method is to learn a map that transforms a simple, well-understood density such as a Gaussian to the target probability distribution through successive transformations. Let $\textbf{x}$ be a $D$-dimensional real vector and $p(\textbf{x})$ its joint distribution we wish to estimate. We proceed by defining a transformation $f$ from a base distribution $\pi(\textbf{u})$ such that
\begin{equation}
    \textbf{x} = f(\textbf{u}), \quad \textbf{u}\sim \pi(\textbf{u}).
    \label{eq:normalizing_flow_sampling}
\end{equation}
We further require that $f$ is a \textit{diffeomorphism}, i.e., that $f$ invertible and that $f$ and $f^{-1}$ are differentiable, which restricts $\textbf{u}$ to be $D$-dimensional as well \cite{Milnor_1997}. The distribution over $\textbf{x}$ can therefore be obtained by the change of variables \cite{Rudin_2006, Bogachev_2007}
\begin{equation}
    p(\textbf{x}) = \pi\left(f^{-1}(\textbf{x})\right) \left| \text{det}\left(\frac{\partial f^{-1}}{\partial \textbf{x}}\right) \right|.
    \label{eq:normalizing_flow_density}
\end{equation}

The composable property of diffeomorphisms, i.e., the fact that for any two diffeomorphisms $f_1$ and $f_2$ the composition $f_1 \circ f_2$ is also diffeomorphic, guarantees that Eq. \eqref{eq:normalizing_flow_density} is still valid if we build a complex transformation from a series of simple transformations $f=f_T\circ\cdots\circ f_1$, where $f_t$ transforms $\textbf{z}_{t-1}$ into $\textbf{z}_{t}$, and setting $\textbf{z}_{0}=\textbf{u}$ and $\textbf{z}_{T}=\textbf{x}$. \textit{Flow} therefore refers to these successive transformations applied on the samples from $\pi(\textbf{u})$ to progressively deform them into the ones of $p(\textbf{x})$, while \textit{normalizing} refers to the inverse flow $f^{-1}_T\circ\cdots\circ f^{-1}_1$ that maps the samples from the distribution $p(\textbf{x})$ into the ones of $\pi(\textbf{u})$, which is often chosen to be a multivariate \textit{normal} \cite{Papamakarios_2019}.

The fact that normalizing flows provide explicit densities (via Eq. \ref{eq:normalizing_flow_density}) in addition to sampling (via Eq. \ref{eq:normalizing_flow_sampling}) is in contrast to other generative models like Variational Autoencoders (VAEs) \cite{Kingma_2013} and Generative Adversarial Networks (GANs) \cite{Goodfellow_2014}, where usually only sampling is possible. In practice, $f$ (or $f^{-1}$) is implemented as a neural network parametrized by $\phi$ (e.g., its weights and biases), which is optimized to learn the mapping between the distribution parameters and a given simulated dataset. The model should have tractable inverse and Jacobian determinant; in other words, it is often required that the inverse is efficient to calculate and that the Jacobian determinant time cost should be at most $\mathcal{O}(D)$.

\begin{figure}
    \centering
    \includegraphics[width=1.0\textwidth]{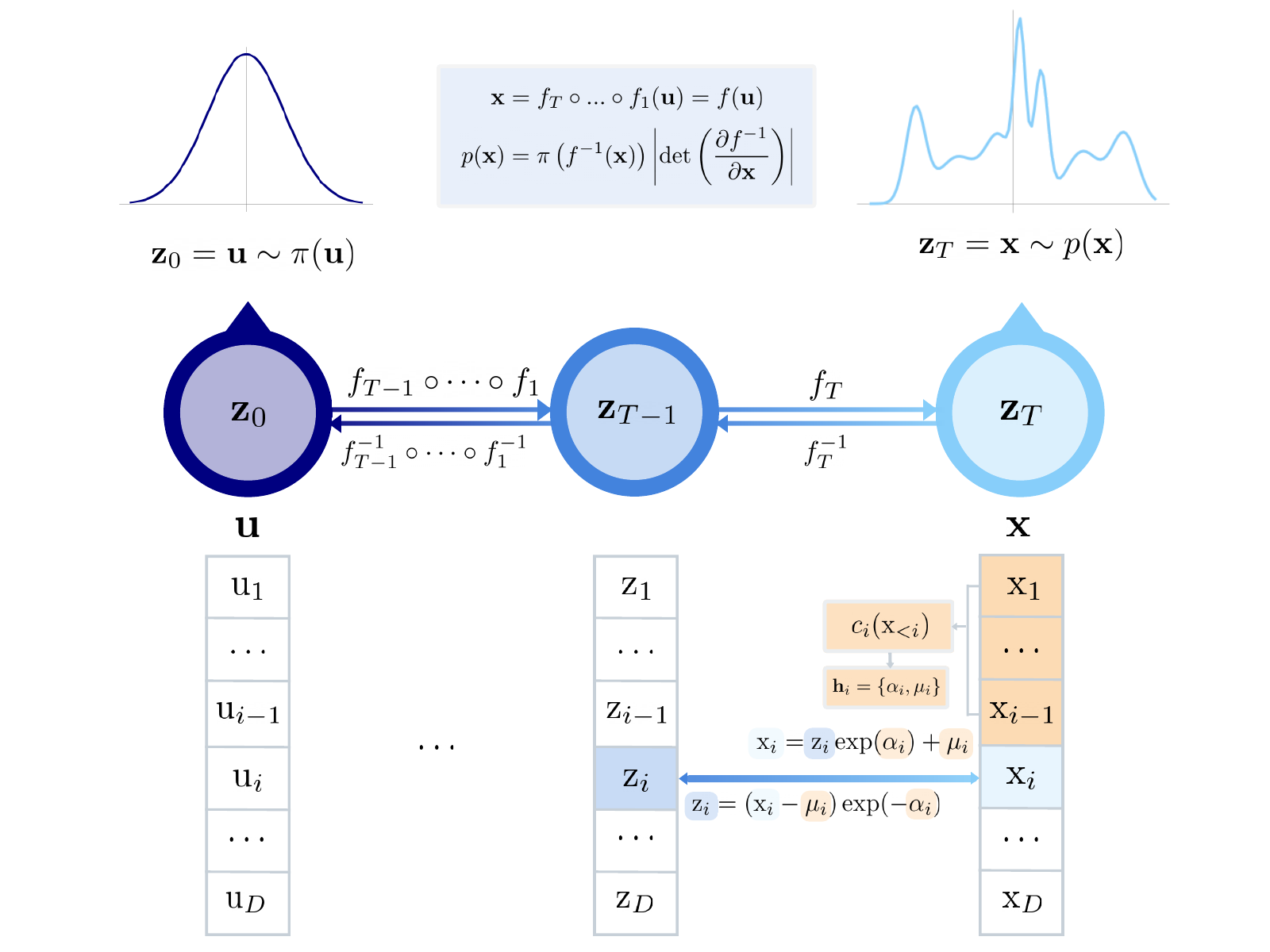}
    \caption{Diagram of how normalizing flows work, with the specific example of Masked Autoregressive Flows. The samples from the vector $\textbf{z}_0=\textbf{u}$, sampled from the simple distribution $\pi(\textbf{u})$, are deformed through the sequence of transformations $f=f_T\circ\cdots\circ f_1$ into those of $\textbf{z}_T=\textbf{x}$, which follow a more complex distribution $p(\textbf{x})$. In the lower panel, we illustrate the conditioner that ``masks out'' the connections between $\text{z}_i$ and $\textbf{h}_{<i}$, as well as the affine functions applied to the vector components.}
    \label{fig:NF}
\end{figure}

\paragraph{Masked Autoregressive Flows.} A simple way to achieve the desirable properties of the model is to use \textit{autoregressive flows}, which restrict $f_t$ to be of the form \cite{Huang_2018}
\begin{equation}
    \text{z}_i' =\tau(\text{z}_i;\textbf{h}_i), \quad \textbf{h}_i = c_i(\textbf{z}_{<i}).
\end{equation}
Here, the output $\text{z}_i'$ is the $i$-th component of a vector $\textbf{z}_t\in\mathbb{R}^D$ for which $\textbf{z}_t=f_t(\textbf{z}_{t-1})$, while $\text{z}_i$ is the $i$-th component of $\textbf{z}_{t-1}\in\mathbb{R}^D$.
The transformer $\tau$ is required to be a strictly monotonic function of $\text{z}_i$, so that the model is invertible. Since $\tau$ is parametrized by $\textbf{h}_i$, the output $\text{z}_i'$ does not depend on $\text{z}_{\geq i}$ due to the \textit{autoregressive} structure of the $i$-th conditioner $c_i$.  As a consequence, the Jacobian is triangular and its determinant can be calculated as the product of its diagonal elements, leading to an $\mathcal{O}(D)$ evaluation of the logarithm of the determinant \cite{Papamakarios_2019}:
\begin{equation}
    \log\left| \text{det}\left(\frac{\partial f}{\partial \textbf{x}}\right) \right| 
    = \log \left| \prod_{i=1}^{D} \frac{\partial \tau}{\partial \text{z}_i}(\text{z}_i;\textbf{h}_i) \right|
    = \sum_{i=1}^{D} \log \left| \frac{\partial \tau}{\partial \text{z}_i}(\text{z}_i;\textbf{h}_i) \right|.
\end{equation}
The implementation of an autoregressive flow then follows by choosing an appropriate transformer $\tau$ and conditioner $c_i$. A common choice for the transformer is to use affine functions, i.e.,
\begin{equation}
    \tau(\text{z}_i;\textbf{h}_i) = \text{z}_i \exp\alpha_i + \beta_i, \quad \textbf{h}_i = \{\alpha_i, \beta_i\},
    \label{eq:affine_flow}
\end{equation}
where we choose to parametrize the scale parameter of this location-scale transformation as $\exp\alpha_i$ to guarantee its invertibility. The determinant of the Jacobian is then simply
\begin{equation}
    \left| \text{det}\left(\frac{\partial f^{-1}}{\partial \textbf{x}}\right) \right| 
    = \exp\left(-\sum_{i=1}^{D}\alpha_i\right).
    \label{eq:affine_Jacobian}
\end{equation}

We focus on Masked Autoregressive Flows (MAFs) \cite{Papamakarios_2017} (see the flowchart \reffig{NF}), a particular case of autoregressive flows which will be used throughout this paper for density estimation, but other alternatives exist such as Inverse Autoregressive Flow (IAF) \cite{Kingma_2016} and Real NVP \cite{Dinh_2016}. MAFs are a special case of \textit{autoregressive models} \cite{Uria_2016}, which use the product rule of probability to decompose densities into 1-dimensional ones as
\begin{equation}
    p(\textbf{x}) = \prod_{i=1}^D p(\text{x}_i|\textbf{x}_{<i}).
\end{equation}
In MAFs, the conditionals are parametrized by Gaussians,
\begin{equation}
    p(\text{x}_i|\textbf{x}_{<i}) = \mathcal{N}(\text{x}_i;\mu_i,(\exp\alpha_i)^2), \quad \textbf{h}_i = \{\alpha_i, \mu_i\},
\end{equation}
which can be equivalently written as an affine function (Eq. \ref{eq:affine_flow})
\begin{equation}
    \text{x}_i = \text{u}_i \exp\alpha_i + \mu_i, \quad \text{u}_i = \mathcal{N}(0,1),
\end{equation}
where the base distribution is normal. Given that the model is easily invertible, with $\text{u}_i=(\text{x}_i-\mu_i)\exp(-\alpha_i)$, and it has a triangular Jacobian of the form of Eq. \eqref{eq:affine_Jacobian}, autoregressive models can also be interpreted as a normalizing flow \cite{Kingma_2016}, as Eqs. \eqref{eq:normalizing_flow_sampling} and \eqref{eq:normalizing_flow_density} can be evaluated. 

The MAF conditioner $c_i$ in turn is implemented with \textit{masked conditioners} as in Masked Autoencoder for Distribution Estimation (MADE) \cite{Germain_2015}, where a single feedforward network outputs $\textbf{h}$ in one pass given the input $\textbf{z}$. To preserve the autoregressive property of the conditioner, each weight matrix element is multiplied by a binary mask, where the connections between $\text{z}_i$ and $\textbf{h}_{<i}$ are ``masked out'' by essentially multiplying their corresponding weights by zero. 

The interpretation of autoregressive models as normalizing flows enables MAF to stack multiple layers of MADEs with Gaussian conditionals into a deeper flow, where the output of each layer, $\alpha_i$ and $\mu_i$ for all $i$, are used as input for the next one. In the language of the previous section, we will have that a MAF transforms the basis function $\pi(\textbf{u})=\mathcal{N}(\mathbf{0},\mathbf{I})$ to the target density $p(\textbf{x})$ through the composition of transformations $f=f_T\circ...\circ f_1$, where each $f_t$ is implemented by a MADE for $\textbf{z}_{0}=\textbf{u}$, $\textbf{z}_{t}=f_t(\textbf{z}_{t-1})$ and $\textbf{z}_{T}=\textbf{x}$.

As a result of the stacking, the model is still tractable, but more expressive and flexible. For example, it has been shown that a MAF with 5 autoregressive unimodal conditional layers is able to approximate multimodal posteriors \cite{Papamakarios_2017}. Another subtlety is that the estimated densities can depend on the order of the inputs, so MAF avoids this issue by using different input orders for each layer. We discuss below how such models are trained for density estimation.

\section{Simulation-based calibration}
\label{sec:sbc}

When using approximated methods for Bayesian inference, a lot of effort has to be made to analyse whether the obtained posterior is correct. For example, for MCMC one has to check whether the chains have converged, and to this date there exists no method to assure convergence with complete confidence \cite{Hermans_2021}. Regarding SBI algorithms, an important diagnostic is to check whether the posterior is calibrated, i.e., not over- or under-estimating the parameter uncertainties. For scientific reasoning, we are mostly concerned with assuring that the posterior is not superconfident. Simulation-based calibration (SBC) \cite{Talts_2018} arises as a useful tool for this analysis, where it is important to stress that passing this test is only a sufficient condition; that is, if it fails, it is an indication that the training was not successful (for example, if not enough simulations were used), while if it passes there is no guarantee that the posterior is correct. Providing reliable tests for SBI algorithms is an active area of research \cite{Hermans_2021, Zhao_2021, Linhart_2022, Lemos_2023a}. 

Here, we use SBC to analyse the estimated posteriors, and its results can tell us if the posterior is not well-calibrated and indicate possible systematic biases on the inference. The idea behind SBC is that, if we draw samples $\boldsymbol{\theta}\sim p(\boldsymbol{\theta})$ from the prior and then generate a data vector $\mathbf{x}\sim p(\mathbf{x}|\boldsymbol{\theta})$ with the simulator, by drawing samples $\hat{\boldsymbol{\theta}}\sim \hat{p}(\hat{\boldsymbol{\theta}}|\textbf{x})$ from the estimated posterior approximated by the model $q_\phi(\hat{\boldsymbol{\theta}}|\textbf{x})$, we obtain joint distribution $\mu(\boldsymbol{\theta}, \mathbf{x}, \hat{\boldsymbol{\theta}}) = p(\boldsymbol{\theta}) p(\mathbf{x}|\boldsymbol{\theta}) \hat{p}(\hat{\boldsymbol{\theta}}|\textbf{x})$. The marginal distribution of $\hat{\boldsymbol{\theta}}$,
\begin{equation}
\begin{split}
    \mu(\hat{\boldsymbol{\theta}}) &= \int d\mathbf{x} \int d\boldsymbol{\theta} \, p(\boldsymbol{\theta}) p(\mathbf{x}|\boldsymbol{\theta}) \hat{p}(\hat{\boldsymbol{\theta}}|\textbf{x}) \\
    &= \int d\mathbf{x} \int d\boldsymbol{\theta} \, p(\mathbf{x},\boldsymbol{\theta}) \hat{p}(\hat{\boldsymbol{\theta}}|\textbf{x}) \\
    &=\int d\mathbf{x} \, p(\mathbf{x}) \hat{p}(\hat{\boldsymbol{\theta}}|\textbf{x}),
\end{split}
\end{equation}
should be distributed according to the prior if the estimated posterior equals the target distribution; that is, if $\hat{p}(\hat{\boldsymbol{\theta}}|\textbf{x}) = p(\hat{\boldsymbol{\theta}}|\textbf{x})$, then 
\begin{equation}
    \mu(\hat{\boldsymbol{\theta}}) = \int d\mathbf{x} \, p(\mathbf{x},\hat{\boldsymbol{\theta}}) = p(\hat{\boldsymbol{\theta}}).
\end{equation}
Analysing the distribution of $\hat{\boldsymbol{\theta}}$ can therefore be used as a posterior calibration test. In practice, we generate a set of ``observed'' data $\textbf{x}_o^i$, where $\textbf{x}_o^i\sim p(\mathbf{x}|\boldsymbol{\theta}_i)$ is simulated given a particular $\boldsymbol{\theta}_i\sim p(\boldsymbol{\theta})$ drawn from the prior, and then estimate the posterior $\hat{p}(\boldsymbol{\theta}|\textbf{x}_o^i)$ for each of these observations. We then draw a set of posterior samples $\{\hat{\boldsymbol{\theta}}\}_i$ for each of the estimated posteriors and compute the \textit{rank} of the observed data under this set by counting how many of the posterior samples $\{\hat{\boldsymbol{\theta}}\}_i$ fall below the corresponding observed data $\textbf{x}_o^i$. If the rank statistics is not uniformly distributed, then the estimated posterior is not well calibrated.

\section{Tests of the inference}
\label{sec:sec_bias_tests}

\begin{figure}[t]
    \centering
    \includegraphics[width=0.49\textwidth]{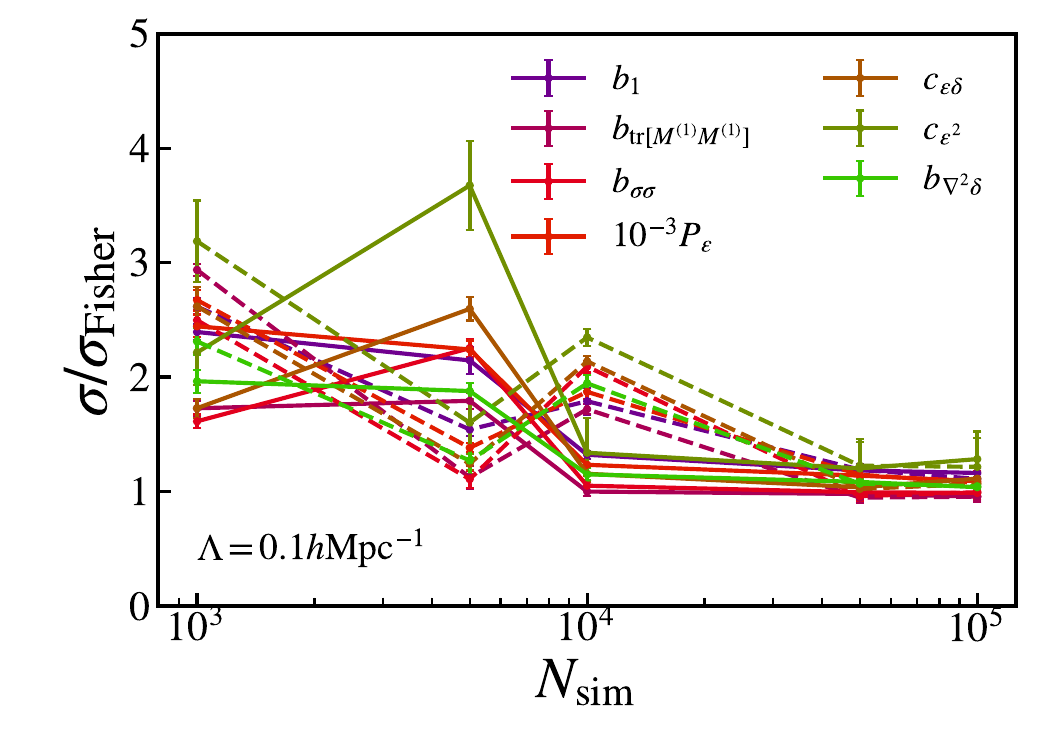}
    \includegraphics[width=0.49\textwidth]{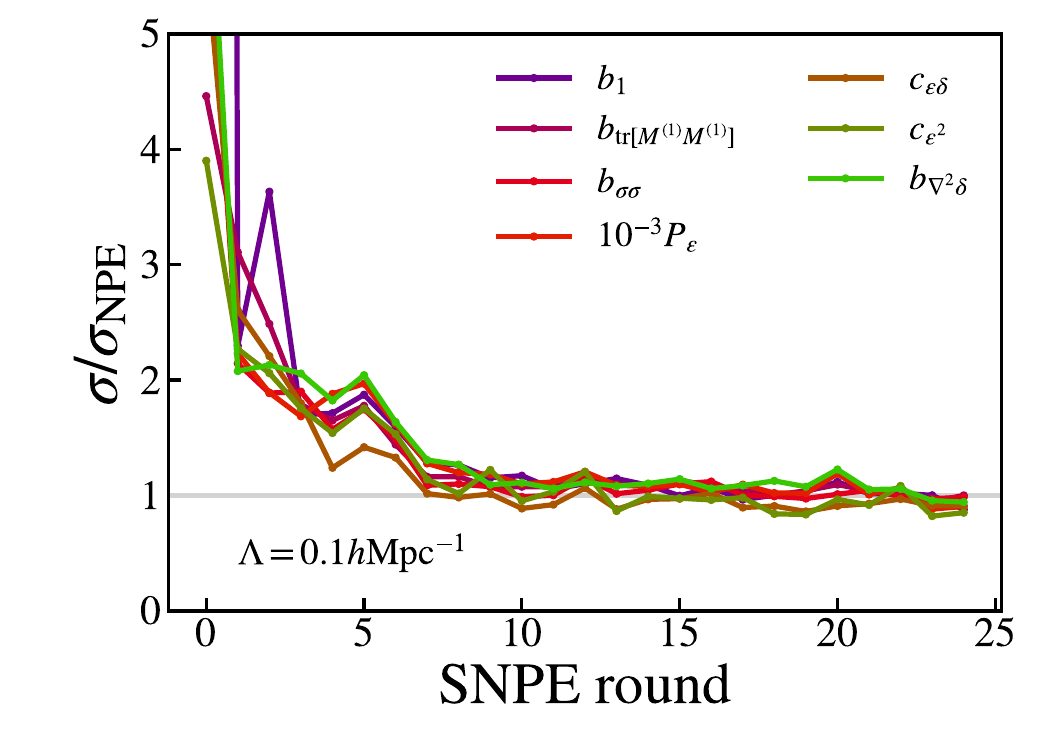}
    \caption{\textit{Left:} Convergence of the standard deviation of the posterior of each parameter with respect to the simulation budget $N_\text{sim}$ normalized by their Fisher prediction. Solid lines denote the case where the data vector corresponds to the raw spectra, while dashed lines display the results using the normalized spectra. \textit{Right:} Standard deviation of the parameters from SNPE starting from the prior of Eq. \eqref{eq:snpe_prior} as a function of its rounds, where $10^3$ simulations are sampled at each round, divided by the NPE standard deviation of each parameter obtained from $N_\text{sim}=10^5$.}
    \label{fig:norm_snpe}
\end{figure}

In this section, we analyse some details of our pipeline focusing on a particular case, namely the one where the cosmological parameter $\alpha$ is fixed and $k_{\text{max}}=\Lambda=0.1h\text{Mpc}^{-1}$. We will be referring to the full case using NPE and $N_\text{sim}=10^5$ unless stated otherwise; note that the following conclusions hold qualitatively the same for the other cases (using a simulation budget $N_\text{sim}$ that guarantees convergence).

\paragraph{Data vector normalization.} It is well known that usually machine learning techniques perform better with normalized values. Since our parameters $\boldsymbol{\theta}$ are already of order one, instead of using the data vector $\textbf{x}$ as the raw power spectrum and bispectrum for density estimation, we also test normalizing them as $P(k)/P_L(k)$ and $B(k_1,k_2,k_3)/[P_L(k_1)P_L(k_2)+\text{2 perm.}]$. However, as we can see in the left panel of Figure \ref{fig:norm_snpe}, the normalization does not lead to faster convergence in this case.

\paragraph{SNPE.} The right hand side of  Figure \ref{fig:norm_snpe} shows the results of SNPE, the sequential version of NPE. As we can see, the standard deviations converge to the ones corresponding to the final NPE posterior, what further confirms our posterior convergence. As aforementioned, although this case uses less simulations for convergence (e.g., after 15 rounds, where convergence seems to be safely reached, one would have used a total of only $N_\text{sim}\sim 10^4$ simulations), performing SBC tests on such non-amortized posteriors is a heavy computational task.

\begin{figure}
    \centering
    \includegraphics[width=0.49\textwidth]{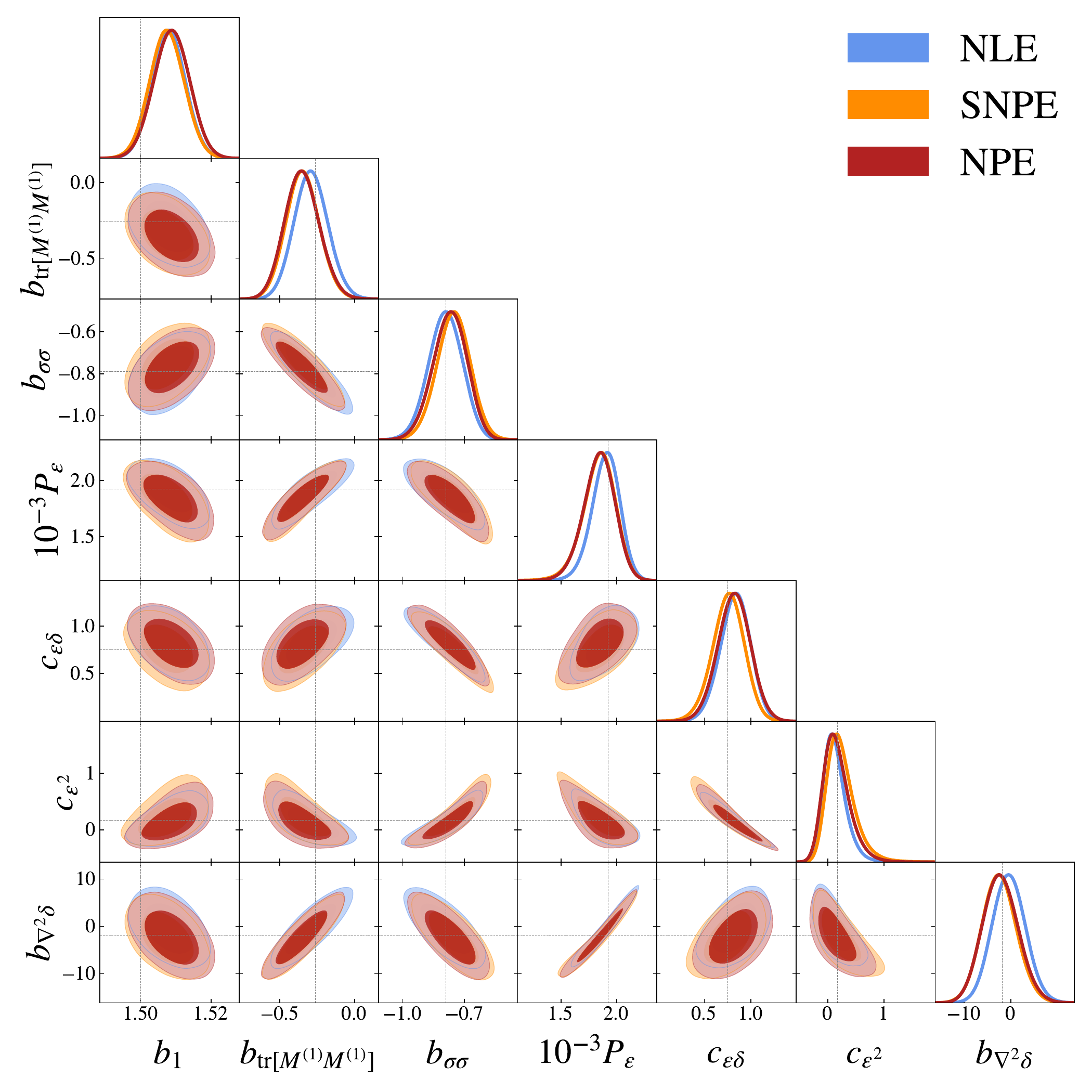}
    \includegraphics[width=0.49\textwidth]{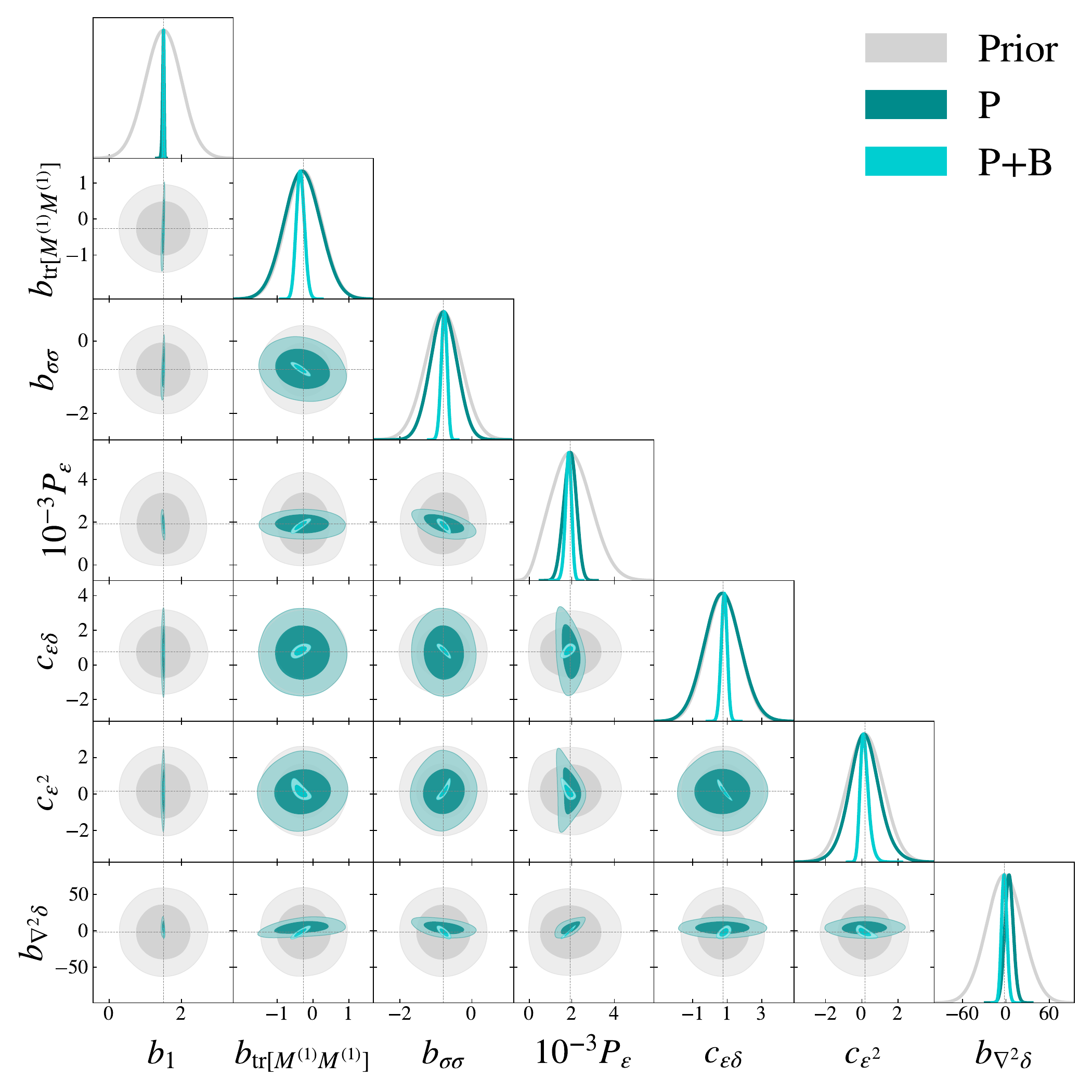}
    \caption{\textit{Left:} posteriors obtained with NLE, NPE and SNPE (at round 22). \textit{Right:} Parameter posterior comparing the prior range in grey with the constraints from power spectrum only (dark cyan) and the power spectrum combined with the bispectrum (cyan).}
    \label{fig:PvsPB}
\end{figure}

\paragraph{Neural density estimation algorithm.} We compare the posteriors obtained from NPE, NLE and SNPE (at round 22) on the left side of Figure \ref{fig:PvsPB}. As expected, all posteriors look very similar. The small deviations could be due to a not sufficiently flexible model or convergence issues. We explore the impact of network architecture below.

\paragraph{Bispectrum constraining power.} We can also compare our analysis with a power-spectrum only analysis, i.e. the case where the data vector consists solely of the power spectrum in the same $k$ bins. We can see on the right hand side of Figure \ref{fig:PvsPB} that, as expected, $b_1$ and $10^{-3} \,P_{\varepsilon}$ are the most well-constrained parameters when considering only the power spectrum. Although the power spectrum does include higher-order (loop) contributions, these have very similar shape and are of limited use in disentangling the second-order bias parameters. As a result, essentially all of these parameters turn out to be prior dominated, i.e. the bispectrum is essential for constraints on higher-order bias and $\sigma_8$.

\begin{figure}
    \centering
    \includegraphics[width=0.49\textwidth]{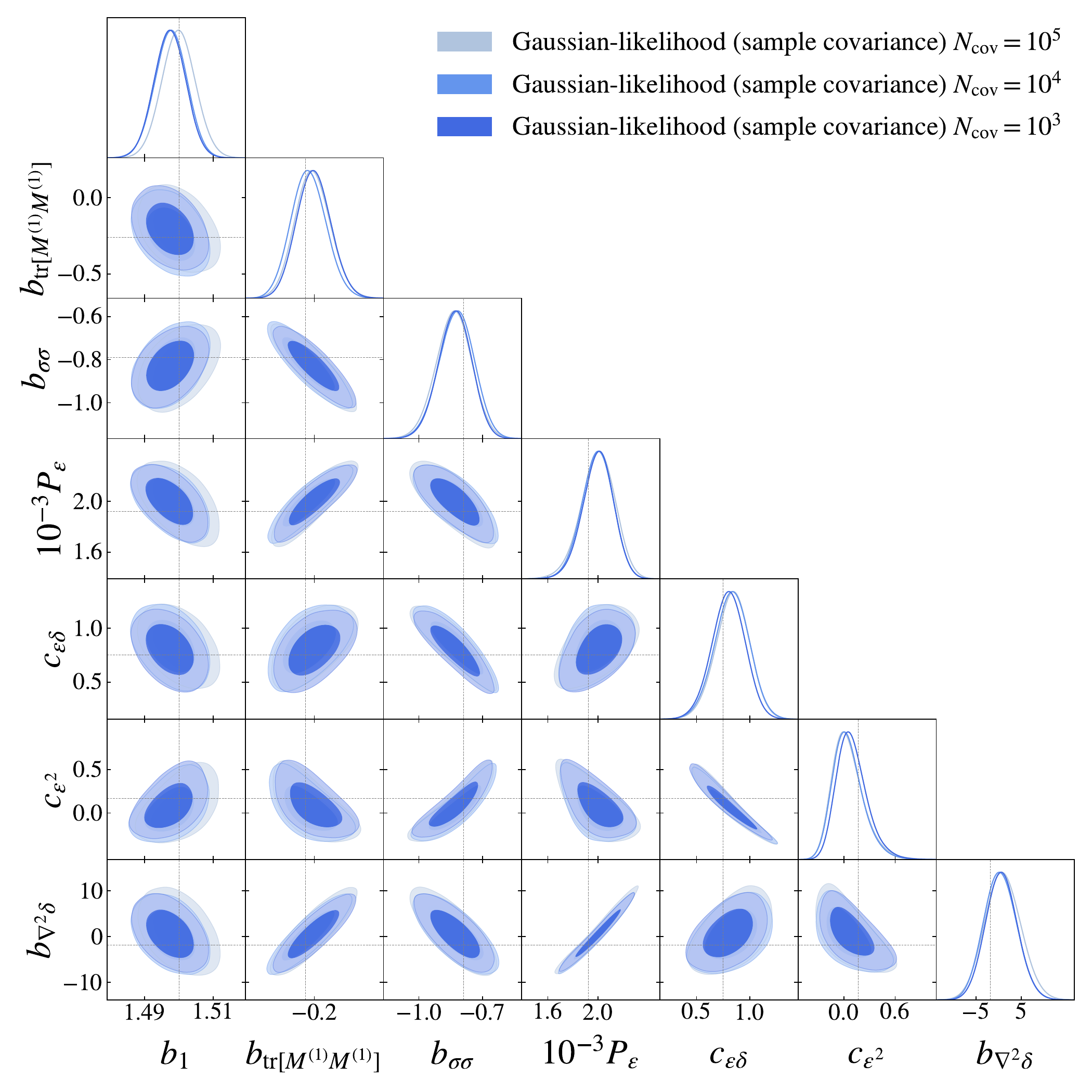}    
    \includegraphics[width=0.49\textwidth]{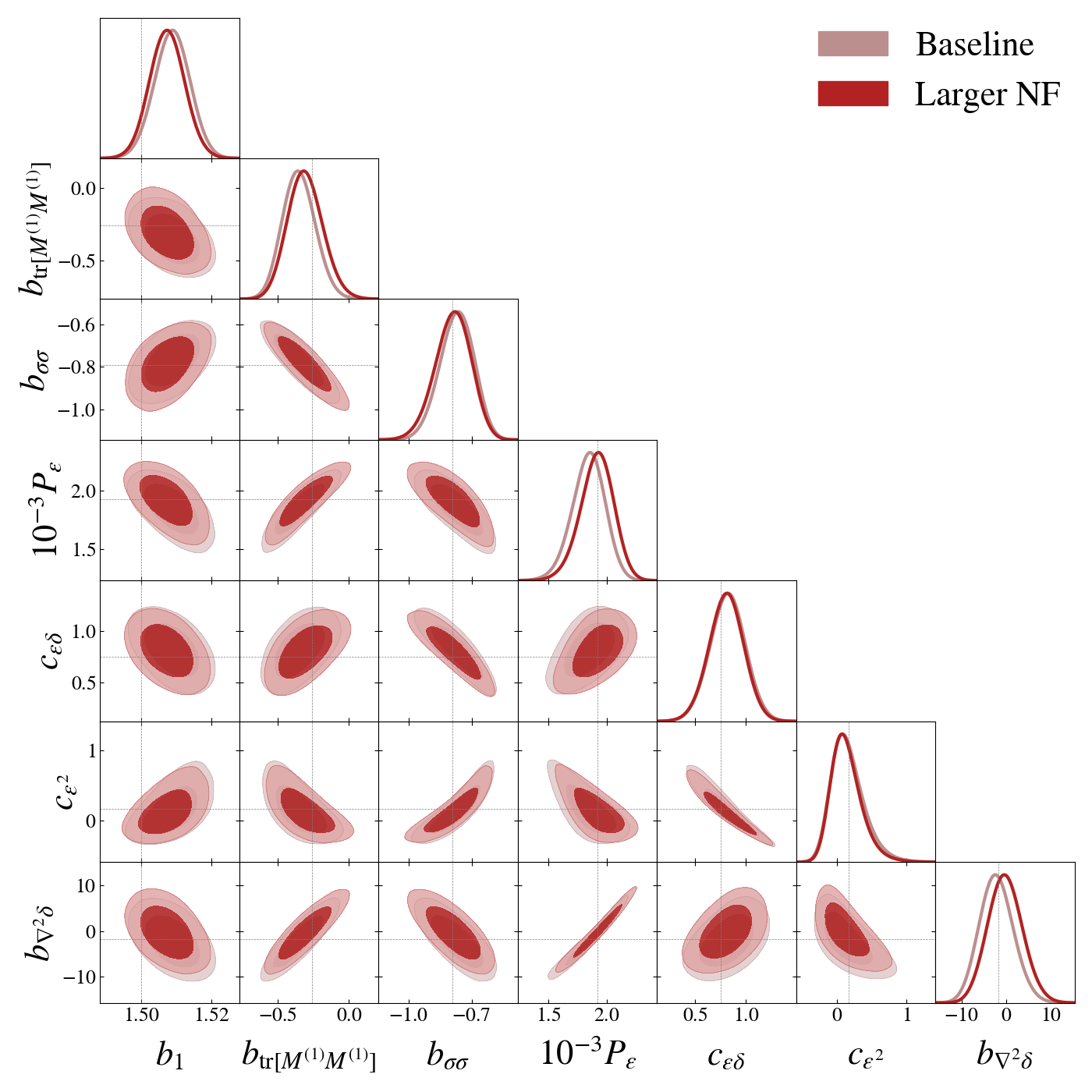}
    \caption{\textit{Left:} impact of the number of simulations for covariance estimation on the final posterior for the Gaussian-likelihood case with sample covariance. \textit{Right:} impact of doubling the number of hidden units and number of transforms.}
    \label{fig:nle_ncov}
\end{figure}

\paragraph{Number of covariance estimates.}  On the left hand side of Figure \ref{fig:nle_ncov} we can see how $N_\text{cov}$, the number of simulations used to estimate the covariance, changes the contours for the Gaussian-likelihood case with sample covariance. As expected, fewer simulations tend to underestimate the errors, although the effect is minor in this case.

\paragraph{Network architecture.} We have also tested how increasing the complexity of the flow can affect the final posterior densities. We show in the right panel of Figure \ref{fig:nle_ncov} that a more complex model, denoted as ``larger NF'' (normalizing flow), where we increased the hidden units from 50 to 100 and the number of transforms from 5 to 10, leads to basically the same posteriors. 

\begin{figure}
    \centering
    \includegraphics[width=0.75\textwidth]{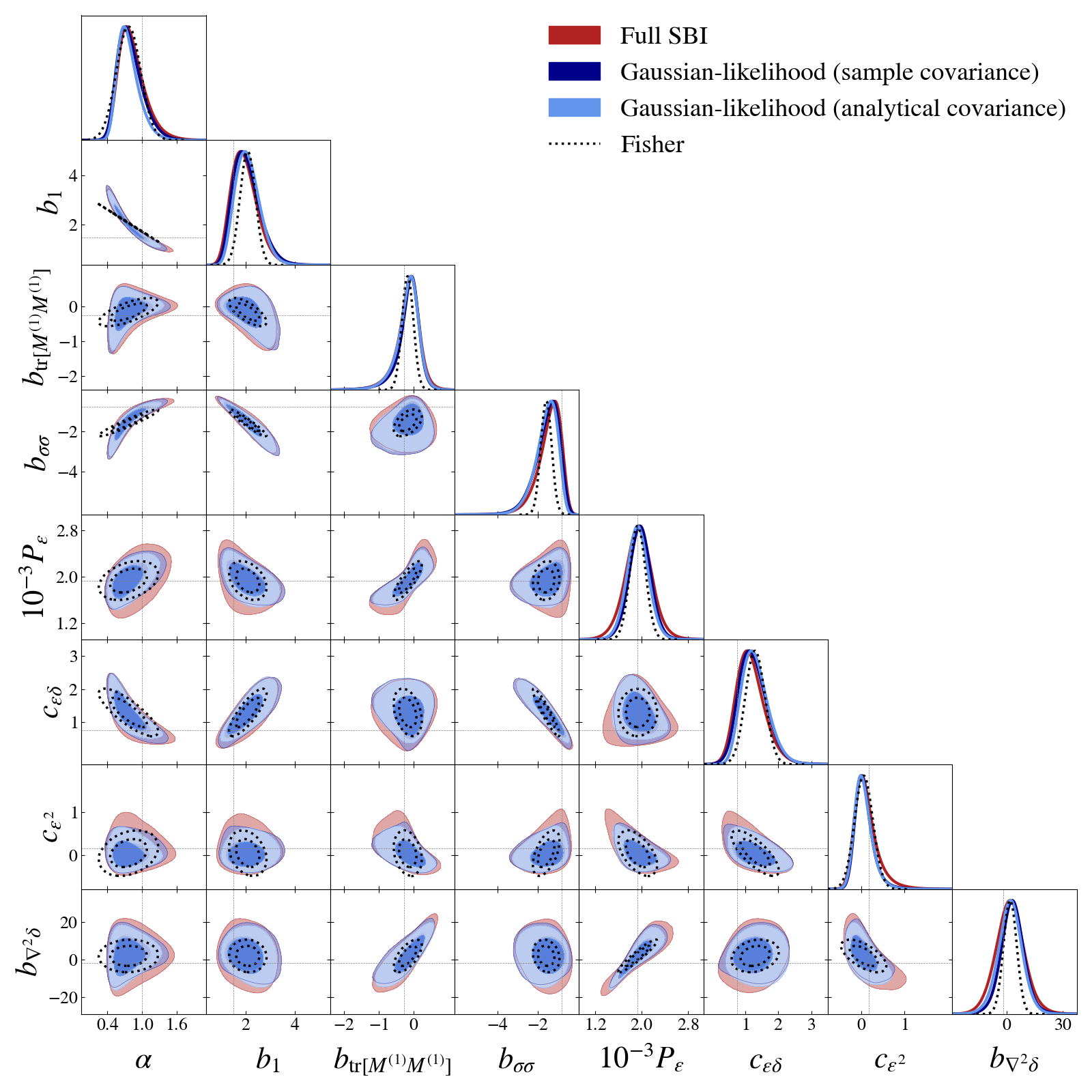}
    \caption{Same as Fig. \ref{fig:alphafree_contours_l01}, but using an ensemble of networks instead of a single posterior estimation.}
    \label{fig:ensemble}
\end{figure}

\paragraph{Ensemble network.} Figure \ref{fig:ensemble} shows the same results as Figure \ref{fig:alphafree_contours_l01}, but with an ensemble of networks instead of a single realization. For that, we independently train 10 different posteriors for the three different cases with the same simulation budget of $N_{\text{sim}}=10^5$, and then sample $10^4$ posterior samples for each of the 10 estimated posteriors. This can give us an estimate of the error associated with the posterior estimation itself, and as we can see the errors are indeed larger; however, the trends are very similar and further confirm our previous findings.

\bibliographystyle{JHEP}
\bibliography{bibliography}

\end{document}